\documentclass[preprint]{ptephy_v1}

\preprintnumber{XXXX-XXXX} 
\usepackage{graphicx}	
\usepackage{amsmath}	
\usepackage{amssymb}	
\usepackage{natbib}
\usepackage{hyperref}
\usepackage{subfigure}
\usepackage{romannum}
\usepackage{cancel}
\usepackage{ulem}

\newcommand{\diff}{{\text{d}}}

\usepackage{color}

\begin{document}

\title{
Self-Similar Solutions for Geometrically Thin Accretion Disks with Magnetically Driven Winds: Application to Tidal Disruption Events
}



\author{Mageshwaran Tamilan}
\affil{Department of Astronomy and Space Science, Chungbuk National University, Cheongju 361-763, Republic of Korea
\email{tmageshwaran2013@gmail.com, tmageshwaran@chungbuk.ac.kr}}
\author[1,2]{Kimitake Hayasaki}
\affil{Department of Physical Sciences, Aoyama Gakuin University, Sagamihara 252-5258, Japan}
\author[3]{Takeru K. Suzuki}
\affil{School of Arts and Sciences, The University of Tokyo, 3-8-1, Meguro, Tokyo 153-8902, Japan}


\begin{abstract}%
We analytically derive self-similar solutions for a time-dependent, one-dimensional, magnetically driven accretion-disk-wind model based on the magnetohydrodynamic equations. The model assumes a geometrically thin, gas-pressure-dominated accretion disk and incorporates both magnetic braking and turbulent viscosity through an extended $\alpha$-viscosity prescription in the vertical and radial directions, respectively. The $\alpha$ parameter for the vertical stress is assumed to vary with the disk aspect ratio. We confirm that in the absence of a wind, our self-similar solutions agree with the classical solution of Cannizzo et al. (1990) \cite{1990ApJ...351...38C}, in which the mass accretion rate follows a power-law decay with time as $t^{-19/16}$. This scaling has been widely used as a key indicator of the mass accretion rate in tidal disruption event (TDE) disks. In contrast, when a wind is present, both the mass accretion and mass loss rates decay more steeply than $t^{-19/16}$. Furthermore, we verify that the power-law indices of these rates are consistent with those obtained from the numerical simulations of Tamilan et al. (2024) at late times. In particular, our analytical solution demonstrates that magnetic braking leads to a more rapid decay of the mass accretion rate, mass loss rate, and bolometric luminosity. In the presence of a strong poloidal magnetic field, all three quantities asymptote to $t^{-5/2}$. This steep decay index can serve as a potential observational signature of magnetocentrifugally driven winds with strong poloidal magnetic fields in TDE disks.
\end{abstract}

\subjectindex{E10, E14, E31, E34, E36}

\maketitle

%

%
\section{Introduction} 
\label{sec:intro}
%
The accretion of matter onto black holes (BHs) serves as the primary energy source for active astrophysical systems, such as X-ray binaries (XRBs), active galactic nuclei (AGNs), and tidal disruption events (TDEs). In the standard model of accretion disks, hydrodynamic eddy turbulence acts as a source of viscosity that redistributes angular momentum within the disk \cite{1973A&A....24..337S}. By introducing the $\alpha$-parameter to describe turbulent viscosity, Shakura \& Sunyaev (1973) \cite{1973A&A....24..337S} modeled a steady-state solution for a geometrically thin, optically thick accretion disk \cite{1981ARA&A..19..137P,2002apa..book.....F,2008bhad.book.....K}. 
Later, Cannizzo et al. (1990) \citep{1990ApJ...351...38C} derived a self-similar solution for a geometrically thin,
gas–pressure–dominated $\alpha$–disk in which the total angular momentum of the disk is conserved; the mass–accretion rate decays as $t^{-19/16}$ at late times.
Subsequently, Pringle (1991) \citep{1991MNRAS.248..754P} obtained an alternative self-similar solution
that conserves the total disk mass, giving $\dot{M}\propto t^{-11/14}$.
Both solutions assume the disk extends smoothly to $r=0$ and therefore possess a single length‐scale, a prerequisite for strict self‐similarity. In practical applications one often truncates the solution at the ISCO; in that context, the ``angular‐momentum‐conserving'' and ``mass‐conserving'' branches correspond to the limiting cases of zero and finite torque at the ISCO, respectively, but this truncation formally breaks exact self‐similarity and should be regarded as an approximation.

Balbus (2017) \cite{2017MNRAS.471.4832B} derived the fully relativistic diffusion equation for a time-dependent, geometrically thin $\alpha$-disk in the Kerr metric. Subsequently,  Balbus \& Mummery (2018) \cite{2018MNRAS.481.3348B} numerically solved this equation by imposing a finite stress at the ISCO and adopting a viscosity prescription which is a function of radius $r$ only. They found that the bolometric luminosity evolves as a power law in time, with a decay rate shallower than $t^{-1}$. In a later study, Mummery \& Balbus (2019) \cite{2019MNRAS.489..132M} generalised the calculation to viscosity prescriptions depending on both radius $r$ and surface density $\Sigma$ and explored both zero- and finite-stress boundary conditions. They showed that the late-time bolometric luminosity recovers $t^{-19/16}$ for zero ISCO stress, whereas a finite stress yields the shallower $t^{-11/14}$ behaviour predicted by the truncated self-similar solutions discussed above. Because the ISCO radius provides a fixed length-scale, these numerical results are not strictly self-similar; nevertheless, the exponents coincide with the analytic scalings once the analytical solution is truncated at the ISCO. These findings indicate that including a finite ISCO stress naturally produces luminosity declines flatter than $t^{-1}$, offering a closer match to the X-ray light curves of TDEs such as PTF-10iya, SDSS~J1201, and OGLE16aaa, which exhibit slopes shallower than $t^{-1}$ \cite{2017ApJ...838..149A}.
%

When a magnetic field is present in the disk, the differential rotation of electrically conducting fluids around a central object induces magnetorotational instability (MRI; \citealp{Velikhov1959,1961hhs..book.....C,1991ApJ...376..214B,1998RvMP...70....1B}). This instability drives magnetohydrodynamic (MHD) turbulence, which produces Maxwell and Reynolds stresses responsible for mass accretion in the disk. Balbus \& Papaloizou \cite{1999ApJ...521..650B} showed that the mean flow behavior of MHD turbulence in a disk aligns with the $\alpha$ prescription. Furthermore, MHD turbulent stresses also generate vertical outflows, whose intensity is governed by the magnetic field strength in the disk \cite{2009ApJ...691L..49S,2014ApJ...784..121S,2013ApJ...765..149C,2019ApJ...872..149L}. MRI amplifies weak magnetic fields, forming large-scale channel flows where magnetic pressure becomes comparable to gas pressure. These channel-mode flows eventually break down due to magnetic reconnection, driving sporadic and intermittent disk winds \cite{2009ApJ...691L..49S}. In the presence of strong poloidal magnetic fields rotating azimuthally with the disk, matter is centrifugally accelerated along field lines, leading to a magnetocentrifugally driven wind when the poloidal field at the disk surface is inclined at an angle greater than 60 degrees, allowing gas to escape gravitational and centrifugal barriers \cite{1982MNRAS.199..883B}. However, there is no clear distinction between the stochastic magnetic pressure-driven process and the magnetocentrifugal mechanism, and they are expected to work together cooperatively.
These MHD-driven outflows extract mass, angular momentum, and energy from the disk, thereby altering its structure and emission.

Suzuki et al. (2016) \cite{2016A&A...596A..74S} developed a time-dependent, one-dimensional (1D) model for a geometrically thin accretion disk with magnetically driven winds, focusing on protoplanetary disks. In their model, they described the magnetohydrodynamic (MHD) turbulence stresses in both radial and vertical directions using an extended $\alpha$ parameterization. Later, Tamilan et al. (2024) \cite{2024ApJ...975...94T} studied a similar 1D, time-dependent, geometrically thin disk with magnetically driven winds, but in the context of TDEs. They modeled the time evolution of an initial Gaussian disk with a zero-torque condition at the innermost stable circular orbit (ISCO). Their results showed that the mass accretion rate decreases more steeply than $t^{-19/16}$ at late times, with the power-law index eventually stabilizing, suggesting a self-similar solution. However, when magnetic braking is included, with a non-zero vertical stress parameter $\alpha$, the mass accretion rate steepens further, becoming steeper than $t^{-2}$. This behavior can explain the light curve dips observed in the TDE candidate AT2019qiz, which follow a $t^{-2.54}$ \cite{2020MNRAS.499..482N}. Tamilan et al. (2025) \cite{2025PTEP.2025b3E02T} developed a steady-state solution for a geometrically thin disk with magnetically driven winds in the context of AGNs and XRBs, where the spectral luminosity deviates from the $\nu^{1/3}$ law when a wind is present, and declines with frequency as the wind intensity increases.

Tabone et al. (2022) \cite{2022MNRAS.512.2290T} derived the self-similar solutions for a 1D time-dependent, geometrically thin accretion disk with magnetically driven winds, by assuming that the disk temperature follows $T \propto r^{-3/2+p}$, where $p$ is a constant parameter. In their model, the disk is vertically isothermal and the temperature does not change over time.
They found that the mass accretion rate follows a power-law dependence on time when radial viscosity dominates over magnetic braking in angular momentum transport, whereas it decreases exponentially with time when magnetic braking dominates over radial viscosity. For the $\alpha$-viscosity prescription, $p=1$ is adopted and in this case, the mass accretion rate decreases as $t^{-3/2}$ in the absence of winds, and its temporal evolution steepens in the presence of winds. However, Tabone et al. (2022) did not solve the energy equation, as they focused on protoplanetary disks, where the disk temperature is mainly determined by radiation from the central star. In contrast, both the standard disk model \cite{1973A&A....24..337S} and the time-dependent model \cite{1990ApJ...351...38C} solved the energy equation, in which the disk temperature follows $T \propto r^{-1/2}\Sigma^{2/3}$, leading to a viscosity that depends on both $r$ and $\Sigma$.
This motivates our focus on deriving a self-similar solution for the magnetically driven disk wind model with a more general form of the disk temperature by explicitly solving the energy equation. Such a model is crucial for understanding the long-term evolution of an accretion disk with a disk wind in XRBs, TDEs, and AGNs, and it also provides the temporal evolution of the disk luminosity, which can be directly compared with TDE observations.

In Section \ref{sec:model}, we briefly describe the basic equations of our model. Section \ref{sec:selfsimilar} derives the self-similar solutions for a 1D, geometrically thin, optically thick, axisymmetric disk with magnetically driven winds. It then presents the time evolution of the mass accretion and loss rates, as well as the bolometric luminosity, based on the self-similar solutions. Additionally, it compares the self-similar solutions with the numerical results obtained by \cite{2024ApJ...975...94T}. Section \ref{sec:discussion} primarily discusses the application of our model to TDEs. Finally, our findings are summarized in Section \ref{sec:conclusion}.

%
\section{Basic Equations} 
\label{sec:model}
%

We consider a 1D, geometrically thin, optically thick, axisymmetric disk with a magnetically driven wind. A time-dependent disk wind model was developed by Suzuki et al. (2016) \cite{2016A&A...596A..74S} for protoplanetary disk systems and by Tamilan et al. (2024) \cite{2024ApJ...975...94T} for TDEs, whereas a steady-state disk model was formulated by Tamilan et al. (2025) \cite{2025PTEP.2025b3E02T} for X-ray binaries and AGNs. The basic equations of the present model are identical to those in Tamilan et al. (2024, 2025) \cite{2024ApJ...975...94T, 2025PTEP.2025b3E02T}. We therefore refer readers to Section 2 and Appendix A of Tamilan et al. (2025) \cite{2025PTEP.2025b3E02T} for their detailed derivation. Here, we provide a concise overview of the essential governing equations, namely those for the conservation of mass, angular momentum, and energy, along with key parameters.

The vertically integrated mass conservation equation is given by
\begin{equation}
\label{eq:sigt}
\frac{\partial \Sigma}{\partial t} 
+ 
\frac{1}{r}\frac{\partial}{\partial r}(r \Sigma v_r) + \dot{\Sigma}_{\rm w}  = 0,
\end{equation}
where $\dot{\Sigma}_{\rm w}$ is the vertical mass flux of the wind, $\Sigma = 2 H \rho$ is the surface density in the disk {with the disk's mass density $\rho$}, $H$ is the disk scale height and $v_r$ is the radial velocity. For a disk with Keplerian rotation:
\begin{equation}
\Omega=\sqrt{\frac{GM}{r^3}},
\label{eq:om}
 \end{equation}
where $M$ is the black hole mass and $G$ is the gravitational constant, 
the angular momentum conservation law yields the following equation: 
\begin{equation}
\label{eq:rsvr}
r \Sigma v_r = -\frac{2}{r\Omega} 
\left[\frac{\partial}{\partial r} 
\left( 
\bar{\alpha}_{r\phi} r^2 \Sigma c_s^2 
\right) 
+ 
\frac{\bar{\alpha}_{z\phi}}{2} 
\frac{r^2 \Sigma c_s^2}{H}
\right],
\end{equation} 
where the sound speed of the disk is given by
\begin{equation}\label{eq:cs}
c_s 
= 
\sqrt{\frac{k_{\rm B} T }{ \mu m_p}},
\end{equation} 
$T$ is the mid-plane temperature of the disk, $k_{\rm B}$ is the Boltzmann constant, $m_{p}$ is the proton mass, $\mu$ is the mean molecular weight taken to be ionized solar mean molecular weight of $0.65$, and $\bar{\alpha}_{r\phi}$ and $\bar{\alpha}_{z\phi}$ are introduced as parameters due to the MHD turbulence and disk winds  
(see the detail for Appendix A of \cite{2025PTEP.2025b3E02T}). The MHD energy equation for our model is given by 
\begin{equation}
\label{eq:ene0}
\dot{\Sigma}_{\rm w} 
\frac{r^2 \Omega^2}{2} 
+
Q_{\rm rad} 
= Q_{+},
\end{equation}
where
$Q_{\rm rad}$ is the radiative cooling flux, which is given by
\begin{equation}\label{eq:qrad}
Q_{\rm rad} = \frac{64\sigma T^4}{3\kappa_{\rm es}\Sigma}
\end{equation}
with $\sigma$ as the Stefan-Boltzmann constant and $\kappa_{\rm es}=0.34~{\rm cm^{2}~g^{-1}}$ as the Thomson scattering opacity, and $Q_+$ represents the heating flux due to turbulent viscosity and magnetic braking as
\begin{equation}
\label{eq:qvisf}
Q_{+} = 
\frac{3}{2}
\bar{\alpha}_{r\phi} 
\Omega\,
\Sigma\,c_{s}^2 
+ 
\frac{1}{2} \bar{\alpha}_{z\phi} r \Omega^2\Sigma c_{s}.
\end{equation} 
We also introduce an additional equation to make a closure of a set of basic equations as
\begin{equation}\label{eq:qradf}
Q_{\rm rad} 
= 
\epsilon_{\rm rad}
\, 
Q_+,
\end{equation}
where $\epsilon_{\rm rad}$ means the fraction of the heating flux of the disk that is converted into radiative cooling flux, and is treated as a constant parameter in the range of $0<\epsilon_{\rm rad}\le1$. 
The hydrostatic balance of the disk is given by
\begin{equation}\label{eq:thindisk} 
c_{s} = \Omega{H}.
\end{equation}
Following Tamilan et al. (2025) \cite{2025PTEP.2025b3E02T}, we assume the magnetic braking parameter to be
\begin{equation}\label{eq:azphi}
\bar{\alpha}_{z\phi} = \bar{\alpha}_{r\phi} \psi \frac{H}{r},
\end{equation}
where $\bar{\alpha}_{r\phi}$ is a constant, typically ranging from 0.01 to 0.1, and $\psi$ is a dimensionless free parameter. This formulation implies that $\bar{\alpha}_{z\phi}$ varies with both radius and time, through its dependence on the disk aspect ratio $H/r$.
A smaller $\psi$, corresponding to lower vertical stress, indicates weaker poloidal fields. In this case, the MRI amplifies the magnetic field, and winds are driven by magnetic pressure and reconnection. In contrast, a higher value of $\psi$ leads to greater vertical stress, which is associated with stronger poloidal fields and more efficient removal of angular momentum from the plasma in the disk. The gas either accretes onto the BH or becomes trapped in the poloidal field. The centrifugal force then accelerates the trapped gas along the poloidal magnetic field, driving the wind. This process is called 
the magnetocentrifugal mechanism. In the limit $\psi \to \infty$, this mechanism operates with maximum efficiency.

Substituting equations~(\ref{eq:rsvr}), (\ref{eq:thindisk}), and (\ref{eq:azphi}) into equation~(\ref{eq:sigt}) yields 
\begin{equation}\label{eq:sdevo3}
\frac{\partial \Sigma}{\partial t} - \frac{2}{r}\frac{\partial}{\partial r}
\left[
\frac{1}{r\Omega}
\left\{
\frac{\partial}{\partial r} 
\left(\bar{\alpha}_{r\phi}  r^2 \Sigma c_{s}^2 \right)
+ 
\frac{1}{2} \psi\bar{\alpha}_{r\phi} r \Sigma c_{s}^2 
\right\}
\right] 
+
\dot\Sigma_{\rm w}=0,
\end{equation}
where $\dot\Sigma_{\rm w}$ is obtained by equating equation~(\ref{eq:ene0}) to equation~(\ref{eq:qvisf}) with equations~(\ref{eq:qradf}), (\ref{eq:thindisk}), and (\ref{eq:azphi}) as
\begin{eqnarray}
\dot{\Sigma}_{\rm w} = 3\left(1-\epsilon_{\rm rad}\right)\bar{\alpha}_{r\phi}\left(1+\frac{\psi}{3}\right)\frac{c_{s}^2 \Sigma}{r^2\Omega},
\label{eq:sigdotw}
\end{eqnarray}
and $c_s^2$ is given by substituting equations~(\ref{eq:qrad}) and (\ref{eq:qvisf}) into equation~(\ref{eq:qradf}) with equations~(\ref{eq:cs}), 
(\ref{eq:thindisk}), and (\ref{eq:azphi}) as
\begin{eqnarray}
c_s^2 = 
\left(
\frac{9}{128}\frac{\kappa_{\rm es}}{ \sigma} 
\right)^{1/3}
\left(\frac{k_{\rm B}}{\mu m_p}\right)^{4/3} 
\epsilon_{\rm rad}^{1/3} 
\bar{\alpha}_{r\phi} ^{1/3}
\left(1+ \frac{\psi}{3} \right)^{1/3}
\Omega^{1/3} \Sigma^{2/3}.
\label{eq:cs6}
\end{eqnarray}
Equations (\ref{eq:sdevo3})-(\ref{eq:cs6}) with equation~(\ref{eq:om}) give the time evolution of the surface density of our disk-wind model. By solving these equations, we obtain the self-similar solutions described in the following sections.

%
\section{Self-similar solutions}
\label{sec:selfsimilar}
%

In this section, we derive self-similar solutions for the basic equations.
%
%
Equation~(\ref{eq:sdevo3}) can be expressed as
\begin{eqnarray}
\frac{\partial \Sigma}{\partial t} 
= 
2
\Lambda
\frac{1}{r}\frac{\partial}{\partial r}
\left[\sqrt{r}\frac{\partial}{\partial r}\left(r^{3/2}\Sigma^{5/3}\right)\right]
+
\psi
\Lambda
\frac{1}{r}
\frac{\partial}{\partial r}
\left(
r \Sigma^{5/3}
\right) 
- 3(1-\epsilon_{\rm rad})
\left(1+\frac{\psi}{3}\right)
\Lambda 
\frac{\Sigma^{5/3}}{r},
\label{eq:sdevo4}
\end{eqnarray}
where 
\begin{equation}
\Lambda
=
 \left(\frac{9\kappa_{\rm es}}{128 \sigma}\right)^{1/3} \left(\frac{k_B}{\mu m_p}\right)^{4/3}
\frac{1}{(G M)^{1/3}}
\bar{\alpha}_{r\phi}^{4/3}
\epsilon_{\rm rad}^{1/3}\left(1+\frac{\psi}{3}\right)^{1/3}
\label{eq:blambda}
\end{equation}
and equation~(\ref{eq:cs6}) was used for the derivation. 
%
%
%
In order to obtain the self-similar solutions, we assume the self-similar form of surface density and self-similar variables as
\begin{eqnarray}
\Sigma 
&=& 
\Sigma_0 \tau^{-\beta}\, F(\xi), 
\label{eq:sssd0}
\\
\xi 
&=& 
\sqrt{\frac{r}{r_0}}\tau^{-\omega}, 
\label{eq:xi} 
\\
\tau 
&=& 
\frac{t}{t_0},
\label{eq:tau}
\end{eqnarray}
respectively, where $F(\xi)$ is the self-similar function, $\Sigma_0$, $r_0$ and $t_0$ are normalization constants, and $\beta$ and $\omega$ represent the power-law indices of the constants.
After some manipulations using equations (\ref{eq:sssd0})-(\ref{eq:tau}), equation~(\ref{eq:sdevo4}) becomes
\begin{eqnarray}
-
\frac{4}{3}
\omega
\frac{d}{d\xi}
\Bigr(
\xi^{\frac{\beta}{\omega}}F
\Bigr)
&=&
\frac{d}{d\xi}
\biggr[
\xi^{\frac{\beta}{\omega}-4}
\frac{d}{d\xi}
\biggr(
\xi^3
F^{5/3}
\biggr)
\biggr]
+
\left(
\psi
-
\frac{\beta}{\omega}
+4
\right)
\frac{d}{d\xi}
\biggr(
\xi^{\frac{\beta}{\omega}-2}
F^{5/3}
\biggr)
\nonumber \\
&-&
\biggr[
\left(
\psi
-
\frac{\beta}{\omega}
+5
\right)
\left(
\frac{\beta}{\omega}-4
\right)
+
6(1-\epsilon_{\rm rad})
\left(1+\frac{\psi}{3}\right)
\biggr]
\xi^{\frac{\beta}{\omega}-3}
F^{5/3},
\label{eq:sdevo5}
\end{eqnarray}
where we adopted for the derivation the following two equations:
\begin{equation}
\frac{2}{3}\beta+2\omega-1 = 0,
\label{eq:ssp1}
\end{equation} 
because of the time independence of $F(\xi)$ and 
\begin{equation}
\frac{2}{3}\Lambda
\frac{\Sigma_0^{2/3} t_0}{r_0}=1,
\label{eq:ssc1}
\end{equation} 
which represents the self-similar condition that the viscous timescale equals to $t_0$. Next, we integrate equation~(\ref{eq:sdevo5}) by adopting the following equation: 
\begin{eqnarray}
\left(
\psi+5
-\frac{\beta}{\omega}
\right)
\left(\frac{\beta}{\omega}-4\right)
+
6(1-\epsilon_{\rm rad})\left(1+\frac{\psi}{3}\right) = 0
\label{eq:ssp2}
\end{eqnarray}
with the boundary condition that $F(\xi) = 0$ at $\xi_{\rm out}$, where $\xi_{\rm out}$ denotes $\xi$ at the outer edge of the disk and $\xi_{\rm out}=1$ is assumed. This integration yields
\begin{eqnarray}
 \frac{d F}{d \xi}
 +
 \frac{3}{5}
 \left(
 \psi
 +
 7
 -
 \frac{\beta}{\omega}
 \right)
\frac{ F}{\xi}
+
\frac{4}{5}
\omega
\xi
F^{1/3}
=0.
\end{eqnarray}
Assuming the solution for the above equation as $F(\xi)=C_1(\xi)\xi^{-(3/5)(\psi+7-\beta/\omega)}$, we get 
\begin{eqnarray}
C_1(\xi)
=
\left[
\frac{4}{15}
\omega
\left(
1+
\frac{1}{5}
\left[
\psi + 7
-\frac{\beta}{\omega}
\right]
\right)^{-1}
\right]^{3/2}
\left[
1
-
\xi^{
2 
+ 
\frac{2}{5}
\left(
\psi + 7
-\frac{\beta}{\omega}
\right)
}
\right]^{3/2}
\end{eqnarray}
so as to satisfy the same boundary condition as above. Note that the boundary condition gives the outer radius of the disk as $r_{\rm out} = r_0 \tau^{2\omega} \xi_{\rm out}^2 = r_0 \tau^{2\omega}$, indicating that the outer radius of the disk increases with time if $\omega > 0$. The increase in $r_{\rm out}$ is caused by the outward transport of angular momentum.
Substituting $F(\xi)$ into equation~(\ref{eq:sssd0}) finally gives the self-similar solution of the surface density as
\begin{eqnarray}
\Sigma
&=&
\Sigma_0
\tau^{-\beta}
\left[
\frac{4}{15}
\omega
\left(
1+
\frac{1}{5}
\left[
\psi + 7
-\frac{\beta}{\omega}
\right]
\right)^{-1}
\right]^{3/2}
\left[
1
-
\xi^{
2 
+ 
\frac{2}{5}
\left(
\psi + 7
-\frac{\beta}{\omega}
\right)
}
\right]^{3/2} \nonumber \\
&&
\times \, \xi^{-(3/5)(\psi+7-\beta/\omega)}.
\label{eq:sdsol}
\end{eqnarray}

\begin{figure}
\centering
\subfigure[]{\includegraphics[scale = 0.635]{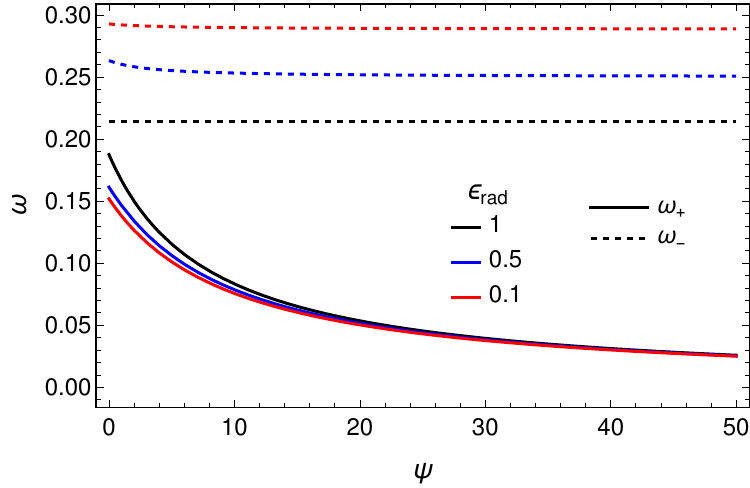}}
\subfigure[]{\includegraphics[scale = 0.635]{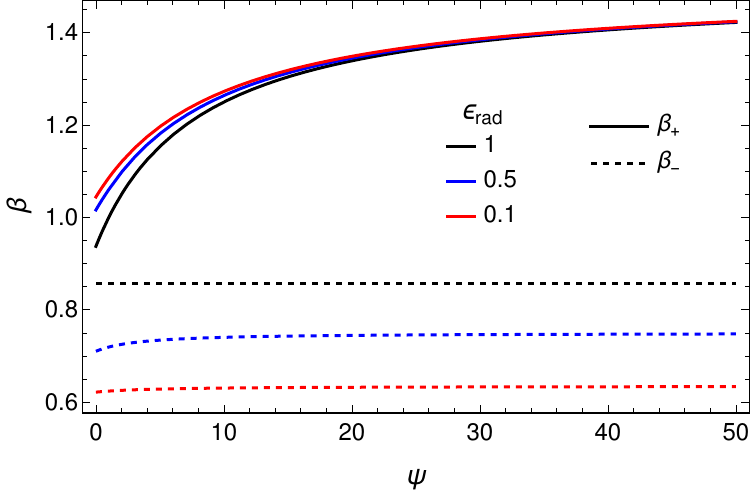}}
\caption{
Dependence of the self-similar power indices $\omega$ and $\beta$ on $\psi$ for the three different values of $\epsilon_{\rm rad}$. Panel (a) shows $\psi$-dependence of $\omega$, while panel (b) shows that of $\beta$. In both panels, the solid and dashed lines denotes $(\beta_{+},\omega_{+})$ and $(\beta_{-},\omega_{-})$, respectively. 
The different color represents the different values of $\epsilon_{\rm rad}$.
}
\label{fig:omegabeta} 
\end{figure}

Equations~(\ref{eq:ssp1}) and (\ref{eq:ssp2}) gives the following quadratic equation for $\omega$ as
\begin{eqnarray}
4\left[
6(1-\epsilon_{\rm rad})
\left(
1+\frac{\psi}{3}
\right)
-
7(\psi+8)
\right]
\omega^2
+
6(\psi+15)\omega
-9=0.
\end{eqnarray}
Solving the quadratic equation gives 
\begin{eqnarray}
\omega
&\equiv& \omega_{\pm} = 
\frac{3}
{
\psi+15
\pm
\sqrt{(\psi+1)^2+8(1-\epsilon_{\rm rad})\left(\psi+3\right)}
} 
\label{eq:omega}
\end{eqnarray}
where we set that $\omega=\omega_{+}$ for the positive signature and that $\omega=\omega_{-}$ for the negative signature.
Substituting equation~(\ref{eq:omega}) into equation~(\ref{eq:ssp1}) gives
\begin{eqnarray}
\beta
&\equiv& \beta_{\pm}
=
\frac{3}{2}
\left[
\frac{
(\psi+9)
\pm
\sqrt{(\psi+1)^2+8(1-\epsilon_{\rm rad})\left(\psi+3\right)}
}
{
\psi+15
\pm
\sqrt{(\psi+1)^2+8(1-\epsilon_{\rm rad})\left(\psi+3\right)}
}
\right]
\label{eq:beta}
\end{eqnarray}
where we also set that $\beta=\beta_{+}$ for the positive signature and that $\beta=\beta_{-}$ for the negative signature. Figure~\ref{fig:omegabeta} confirms that 
$\omega >0$ and $\beta>0$ hold for any value of $\epsilon_{\rm rad}$ within a reasonable range of $\psi$. 

Two solutions of $\omega$ in equation~(\ref{eq:omega}) demonstrate that there are two physically possible solutions for equation~(\ref{eq:sdsol}).
In fact, considering the case of no wind $(\epsilon_{\rm rad}=1,\psi=0)$, the surface density is given by 
\begin{eqnarray}
\Sigma = \left(\frac{1}{28}\right)^{3/2} \Sigma_0 \tau^{-\frac{15}{16}} \xi^{-\frac{6}{5}} \left(1-\xi^{\frac{14}{5}}\right)^{3/2}
\end{eqnarray}
for $(\beta,\omega)=(\beta_+,\omega_+)=(15/16,3/16)$. 
This solution corresponds to that derived by Cannizzo et al. (1990) \cite{1990ApJ...351...38C} for a geometrically thin disk, obtained under the assumption that the total disk angular momentum remains constant. In their numerical simulations with a finite inner disk radius, this condition is implemented by imposing a zero-stress boundary at the inner edge, $r_{\mathrm{in}}$.
Throughout this paper, we assume $r_{\rm in}$ to be the ISCO radius of a non-spinning black hole, i.e., $r_{\rm in} = 6 r_{\rm g}$, where
\begin{equation} 
r_{\rm g} \sim 1.5 \times 10^{11}{\rm cm}\left(\frac{M}{10^6 M_{\odot}}\right) \end{equation}
is the gravitational radius. On the other hand, for $(\beta,\omega)=(\beta_{-},\omega_{-})=(6/7,3/14)$, we obtain the surface density as 
\begin{eqnarray}
\Sigma 
= 
\left(\frac{1}{28}\right)^{3/2} \Sigma_0 \tau^{-\frac{6}{7}} \xi^{-\frac{9}{5}} \left(1-\xi^{\frac{16}{5}}\right)^{3/2}.
\end{eqnarray}
This solution agrees with the self-similar one derived by Pringle (1991) \cite{1991MNRAS.248..754P} for a geometrically thin disk, obtained under the assumption that the total disk mass remains constant. In numerical simulations with a finite inner disk radius, this condition corresponds to setting the mass accretion rate to zero at $r_{\rm in}$.

%
%

%
\subsection{Disk Mass and Angular Momentum Evolution}
\label{sec:mdjd}
%

The disk mass is given by using equations~(\ref{eq:xi}), (\ref{eq:tau}), and (\ref{eq:sdsol}) as
\begin{eqnarray}
M_{\rm d} 
&=& 
\int_{r_{\rm in}}^{r_{\rm out}} \Sigma 2\pi r\,\diff r
= 4\pi\Sigma_0r_0^2\tau^{-\beta + 4 \omega} 
\int_{\xi_{\rm in}(\tau)}^1 
\xi^3 
F(\xi)
\diff \xi,
\nonumber \\
&=& 
4\pi\Sigma_0r_0^2
\left[
\frac{4}{15}
\omega
\left(
1+
\frac{1}{5}
\left[
\psi + 7
-\frac{\beta}{\omega}
\right]
\right)^{-1}
\right]^{3/2}
G_1(\xi_{\rm in})
\,
\tau^{-\beta + 4 \omega}, 
\label{eq:md}
\end{eqnarray} 
where 
\begin{eqnarray}
G_1(\xi_{\rm in})
&\equiv&
\int_{\xi_{\rm in}}^{1}
\left[
1
-
\xi^{
2 
+ 
\frac{2}{5}
\left(
\psi + 7
-\frac{\beta}{\omega}
\right)
}
\right]^{3/2}
\xi^{
-
\frac{3}{5}
\left(
\psi+2-\frac{\beta}{\omega}
\right)
}
d \xi.
\label{eq:G1t}
\end{eqnarray}
Here, $G_1(\xi_{\rm in})$ is a dimensionless function of $\xi_{\rm in}(\tau) = \sqrt{r_{\rm in}/r_0}~ \tau^{-\omega}$, suggesting that $G_1(\xi_{\rm in})$ can change with time and depending on whether it is $\omega_+$ or $\omega_-$. 
To clarify the asymptotic behaviour of $G_1(\xi_{\rm in})$ as the inner edge tends to zero, we divide the integration interval into an inner part $[\xi_{\rm in},\xi_{\rm tr}]$ and an outer part $[\xi_{\rm tr},1]$, where the arbitrary constant $\xi_{\rm tr}$ satisfies $\xi_{\rm in}< \xi_{\rm tr}<1$. For $\xi\le\xi_{\rm tr}$ we may set  
$1-\xi^{\,2+\frac{2}{5}\left(\psi+7-\frac{\beta}{\omega}\right)}\simeq1$, so the integrand reduces to a power law. This gives
\begin{multline}\label{eq:G1split}
G_1(\xi_{\rm in}) = \frac{5}{-1-3\psi+3\beta/\omega}\left[\xi_{\rm tr}^{(-1-3\psi+3\beta/\omega)/5}-\xi_{\rm in}^{(-1-3\psi+3\beta/\omega)/5}\right]\\ +\int_{\xi_{\rm tr}}^{1}
\left[1-\xi^{2+\frac{2}{5}\left(\psi + 7-\frac{\beta}{\omega}\right)}\right]^{3/2}
\xi^{-\frac{3}{5}\left(\psi+2-\frac{\beta}{\omega}\right)}d \xi.
\end{multline}
The second term is finite because its integrand remains regular for $\xi\in[\xi_{\rm tr},1]$. Consequently, $G_1(\xi_{\rm in})$ approaches a constant as $\xi_{\rm in}\to0$ when $(-1-3\psi+3\beta/\omega)/5\ge0$. If $(-1-3\psi+3\beta/\omega)/5<0$, the first term dominates and $G_1(\xi_{\rm in})\propto\xi_{\rm in}^{\,(-1-3\psi+3\beta/\omega)/5}$, implying a divergence in the limit $\xi_{\rm in}\to0$.

Similarly, the angular momentum of the disk is given by
\begin{eqnarray}
J_{\rm d} 
&=& 
\int_{r_{\rm in}}^{r_{\rm out}} r^2\Omega \Sigma 2\pi r\,\diff r
= 
4\pi {r_{0}^2} \Sigma_0 ( r_0^2 \Omega_0)
\tau^{-\beta + 5 \omega}
\int_{\xi_{\rm in}(\tau)}^1 \xi^4 F(\xi)
d\xi 
\nonumber\\
&=&
4\pi {r_{0}^2} \Sigma_0 ( r_0^2 \Omega_0)
\left[
\frac{4}{15}
\omega
\left(
1+
\frac{1}{5}
\left[
\psi + 7
-\frac{\beta}{\omega}
\right]
\right)^{-1}
\right]^{3/2}
G_2(\xi_{\rm in})
\,
\tau^{-\beta + 5 \omega}, 
\label{eq:jd}
\end{eqnarray}
where
\begin{equation}
\begin{aligned}
G_2(\xi_{\rm in})
&\equiv
\int_{\xi_{\rm in}}^{1}
\left[
1
-
\xi^{
2 
+ 
\frac{2}{5}
\left(
\psi + 7
-\frac{\beta}{\omega}
\right)
}
\right]^{3/2}
\xi^{
-
\frac{1}{5}
\left(
1+3\psi-3\frac{\beta}{\omega}
\right)
}
d \xi \,.
\end{aligned}
\label{eq:G2t}
\end{equation}
As with $G_1(\xi_{\rm in})$, the function $G_2(\xi_{\rm in})$ is dimensionless and depends on time. We can decompose $G_2(\xi_{\rm in})$ in a similar way:
\begin{equation}\label{eq:G2split}
\begin{aligned}
G_2(\xi_{\rm in}) &= \frac{5}{4 - 3\psi + 3\beta/\omega}
\left[
\xi_{\rm tr}^{(4 - 3\psi + 3\beta/\omega)/5}
-
\xi_{\rm in}^{(4 - 3\psi + 3\beta/\omega)/5}
\right] \\
&\quad +
\int_{\xi_{\rm tr}}^1
\left[
1 - \xi^{2 + \frac{2}{5} \left(\psi + 7 - \frac{\beta}{\omega} \right)}
\right]^{3/2}
\xi^{-\frac{1}{5} \left(1 + 3\psi - 3\frac{\beta}{\omega} \right)}
d\xi.
\end{aligned}
\end{equation}
Since the integrand in the second term remains regular throughout the interval $[\xi_{\rm tr}, 1]$, it contributes a finite constant. Therefore, the asymptotic behavior of $G_2(\xi_{\rm in})$ as $\xi_{\rm in} \to 0$ depends solely on the exponent $(4 - 3\psi + 3\beta/\omega)/5$. If this exponent is positive, $G_2(\xi_{\rm in})$ converges to a constant; otherwise, it diverges as a power law.

For the $\omega = \omega_+$ solution branch, we find that both exponents
\begin{eqnarray*}
\frac{-1-3\psi+3\beta/\omega}{5} =&& \frac{3\sqrt{(\psi+1)^2+8(1-\epsilon_{\rm rad})(\psi+3)}-3\psi+25}{10}, \\
\\
\frac{4 - 3\psi + 3\beta/\omega}{5} =&& \frac{3\sqrt{(\psi + 1)^2 + 8(1 - \epsilon_{\rm rad})(\psi + 3)} - 3\psi + 35}{10}
\end{eqnarray*}
becomes positive for all values of $\psi \geq 0$ and $0 < \epsilon_{\rm rad} \leq 1$. This confirms that $G_1(\xi_{\rm in})$ and $G_2(\xi_{\rm in})$ approaches a finite value as $\xi_{\rm in} \to 0$, regardless of the parameter choices. This behavior is clearly demonstrated in panels (a) and (c) of Figure~2: for $\omega = \omega_+$, all curves flatten out and become nearly constant for $\xi_{\rm in} \lesssim 0.1$, particularly when $\psi = 0$ (solid lines). Hence, for the $\omega_+$ solution, both the disk mass and angular momentum follow simple power-law time evolutions: $M_{\rm d} \propto \tau^{-\beta + 4\omega}$ and $J_{\rm d} \propto \tau^{-\beta + 5\omega}$. Assuming that $n_{\rm M} = -\beta+4\omega$ and $n_{\rm J} = -\beta+5\omega$, we calculate them using equations~(\ref{eq:omega}) and (\ref{eq:beta}) as
\begin{eqnarray}
n_{\rm M} 
&=&
 -\frac{3}{2}
\left[
\frac{
\psi+1 + \sqrt{(1+\psi)^2+8(1-\epsilon_{\rm  rad})\left(3+\psi\right)}
}
{\psi+15+ \sqrt{(1+\psi)^2+8(1-\epsilon_{\rm  rad})\left(3+\psi\right)}
}
\right]
,
\label{eq:nm}
\\
n_{\rm J} 
&=&
-\frac{3}{2}
\left[
\frac{
\psi-1 + \sqrt{(1+\psi)^2+8(1-\epsilon_{\rm  rad})\left(3+\psi\right)}
}
{
 \psi+15+ \sqrt{(1+\psi)^2+8(1-\epsilon_{\rm  rad})\left(3+\psi\right)}
 }
 \right].
\label{eq:nj}
\end{eqnarray}
For instance, in the absence of wind ($\epsilon_{\rm rad} = 1$, $\psi = 0$), we obtain $n_{\rm M} = -3/16$ and $n_{\rm J} = 0$, indicating that the disk mass declines slowly while the total angular momentum remains constant. In contrast, for $\epsilon_{\rm rad} = 0.5$ with $\psi = 10$, both indices steepen to $n_{\rm M} = -0.95$ and $n_{\rm J} = -0.87$, reflecting stronger depletion due to enhanced wind activity and magnetic braking.

\begin{figure*}
\centering
\subfigure[]{\includegraphics[scale = 0.66]{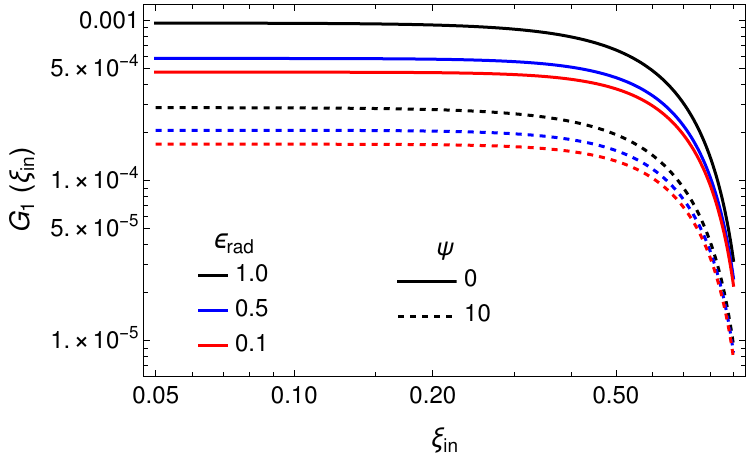}}
\subfigure[]{\includegraphics[scale = 0.62]{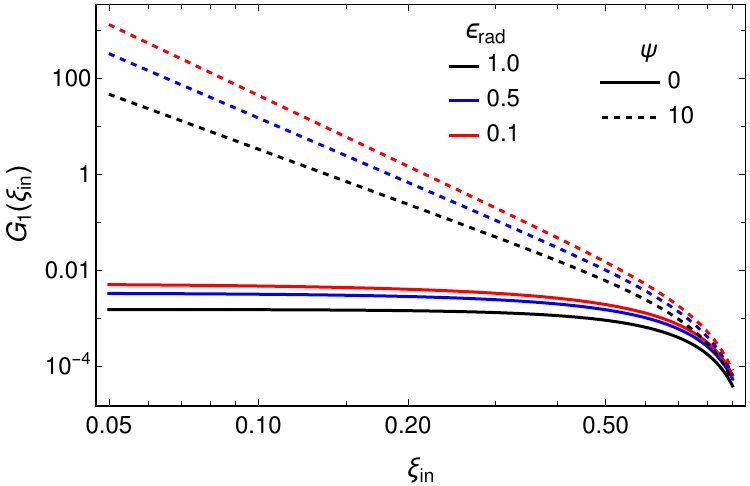}}
\subfigure[]{\includegraphics[scale = 0.66]{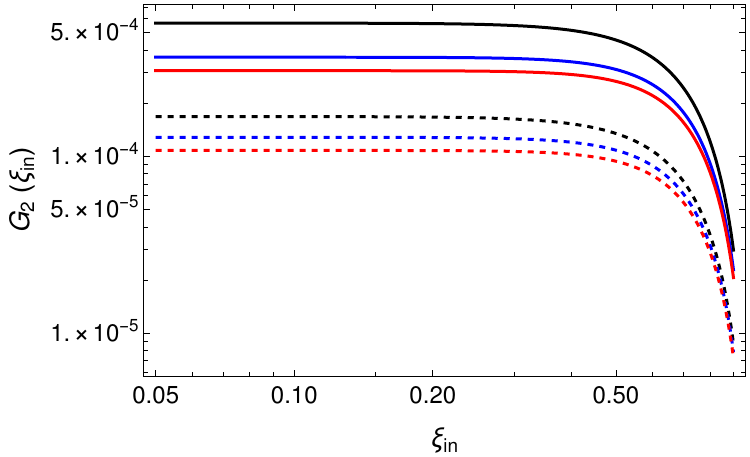}}
\subfigure[]{\includegraphics[scale = 0.62]{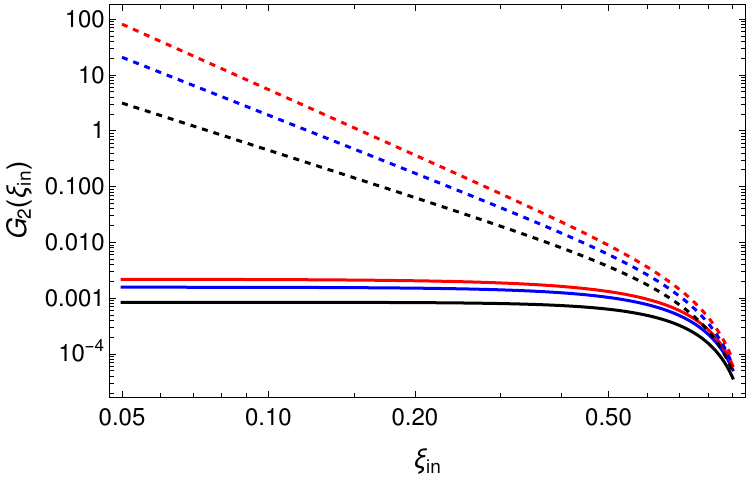}}
\caption{
Dependence of $G_1(\xi_{\rm in})$ and $G_2(\xi_{\rm in})$ on $\xi_{\rm in}$, computed using equations~(\ref{eq:G1t}) and~(\ref{eq:G2t}), respectively, for $\omega_+$ and $\omega_-$. Panels (a) and (c) are for $\omega_+$, while panels (b) and (d) are for $\omega_-$. For all the panels, the different colors represent three different $\epsilon_{\rm rad}$, and the solid and dashed lines show the $\psi=0$ and $\psi=10$ cases, respectively.
}
\label{fig:mdjd}
\end{figure*}

In contrast, the $\omega = \omega_-$ branch behaves quite differently. The exponents
\begin{eqnarray*}
\frac{-1 - 3\psi + 3\beta/\omega}{5} =&& \frac{-3\sqrt{(\psi + 1)^2 + 8(1 - \epsilon_{\rm rad})(\psi + 3)} - 3\psi + 25}{10}, \\
\\
\frac{4 - 3\psi + 3\beta/\omega}{5} =&& \frac{-3\sqrt{(\psi + 1)^2 + 8(1 - \epsilon_{\rm rad})(\psi + 3)} - 3\psi + 35}{10}
\end{eqnarray*}
becomes negative for $\psi > (50 + 27\epsilon_{\rm rad}) / (30 - 9\epsilon_{\rm rad})$ and $\psi > (250 - 54\epsilon_{\rm rad})/(75 - 18\epsilon_{\rm rad})$, respectively. In this regime, $G_1(\xi_{\rm in}) \propto \xi_{\rm in}^{(-1 - 3\psi + 3\beta/\omega)/5}$ and  $G_2(\xi_{\rm in}) \propto \xi_{\rm in}^{(4 - 3\psi + 3\beta/\omega)/5}$ diverges as $\xi_{\rm in} \to 0$. This divergence is clearly shown in panels (b) and (d) of Figure~\ref{fig:mdjd}: for $\psi = 10$ (dashed curves), $G_1(\xi_{\rm in})$ and $G_2(\xi_{\rm in})$ grows sharply at small $\xi_{\rm in}$, especially for lower $\epsilon_{\rm rad}$ values.
%
This implies that for the $\omega_-$ solution with strong magnetic braking, both the disk mass and angular momentum increase with time, leading to an unphysical outcome in the absence of external mass supply. This artificial growth arises because $\xi_{\rm in} \propto \tau^{-\omega_-}$, so power-law divergences in $G_1(\xi_{\rm in})$ and $G_2(\xi_{\rm in})$ translate into increasing $M_{\rm d}$ and $J_{\rm d}$ as shown in Figure \ref{fig:nMnJ_minus} of Appendix~\ref{app:msolution}. Such trends contradict the physical expectation that disk winds deplete, rather than replenish, mass and angular momentum. Therefore, for the remainder of this study, we adopt the $\omega = \omega_+$ solution branch as the physically viable one, unless explicitly stated otherwise.

%
%



%
\subsection{Surface Density in Physical Units}
%
We consider a TDE disk formed by the tidal disruption of a star with mass $m_{\star}$ and radius $r_{\star}$ by a SMBH with mass $M$ \cite{1975Natur.254..295H,1988Natur.333..523R}. At $\tau=1$ corresponding to $t = t_0$, we assume that the outer radius of the disk is the circularization radius, which is typically twice the tidal disruption radius: 
\begin{eqnarray}
r_{\rm t}
&=&
\left(\frac{M}{m_{\star}}\right)^{1/3}r_{\star}
\nonumber \\
&\sim&
7.0 \times 10^{12}~{\rm cm}~\left(\frac{M}{10^6 M_{\odot}}\right)^{1/3}\left(\frac{m_{\star}}{M_{\odot}}\right)^{-1/3}\left(\frac{r_{\star}}{R_{\odot}}\right),
\label{eq:rt}
\end{eqnarray}
i.e., $r_{\rm out} = 2r_{\rm t}$. Also, the total disk mass is assumed to be a half of the stellar mass: $m_{\star}/2$ at $\tau=1$. Given $r_{0} = r_{\rm out}$ and $M_{d}=m_\star/2$ at $\tau = 1$, $\Sigma_0$ is estimated by using equation (\ref{eq:md}) as
\begin{eqnarray}
\Sigma_0 
&=&  \frac{m_{\star}^{5/3}}{32 \pi M^{2/3} r_{\star}^2} \left[\frac{4}{15}\omega\left(1+\frac{1}{5}\left[\psi + 7-\frac{\beta}{\omega}\right]\right)^{-1}\right]^{-3/2}
\biggr/
G_{1}
\left(
\sqrt{\frac{r_{\rm in}}{r_{\rm out}}}
\right)
\nonumber \\
&\sim& 
4.1 \times 10^{5}~{\rm g~cm^{-2}}
\left(\frac{M}{10^6 M_{\odot}}\right)^{-2/3} \left(\frac{m_{\star}}{M_{\odot}}\right)^{5/3}
\left(\frac{r_{\star}}{R_{\odot}}\right)^{-2}  
\nonumber \\
&\times&
\left[\frac{4}{15}\omega\left(1+\frac{1}{5}\left[\psi + 7-\frac{\beta}{\omega}\right]\right)^{-1}\right]^{-3/2}
\biggr/
G_{1}
\left(
\sqrt{\frac{r_{\rm in}}{r_{\rm out}}}
\right).
\label{eq:sigma0}
\end{eqnarray}

\begin{figure*}
\centering
\subfigure[]{\includegraphics[scale = 0.46]{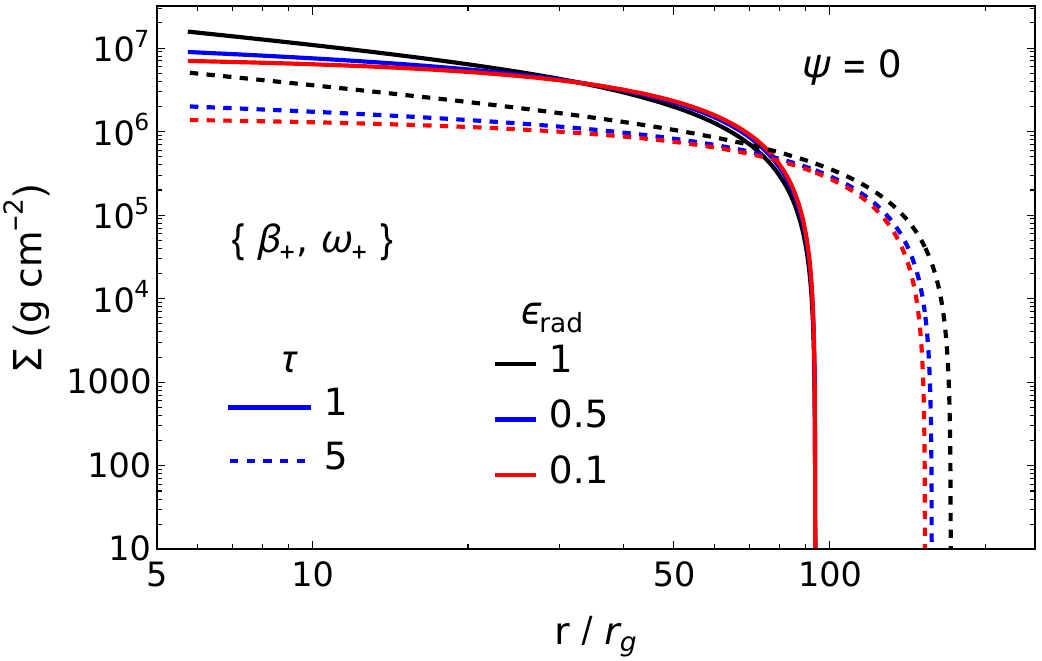}}
\subfigure[]{\includegraphics[scale = 0.46]{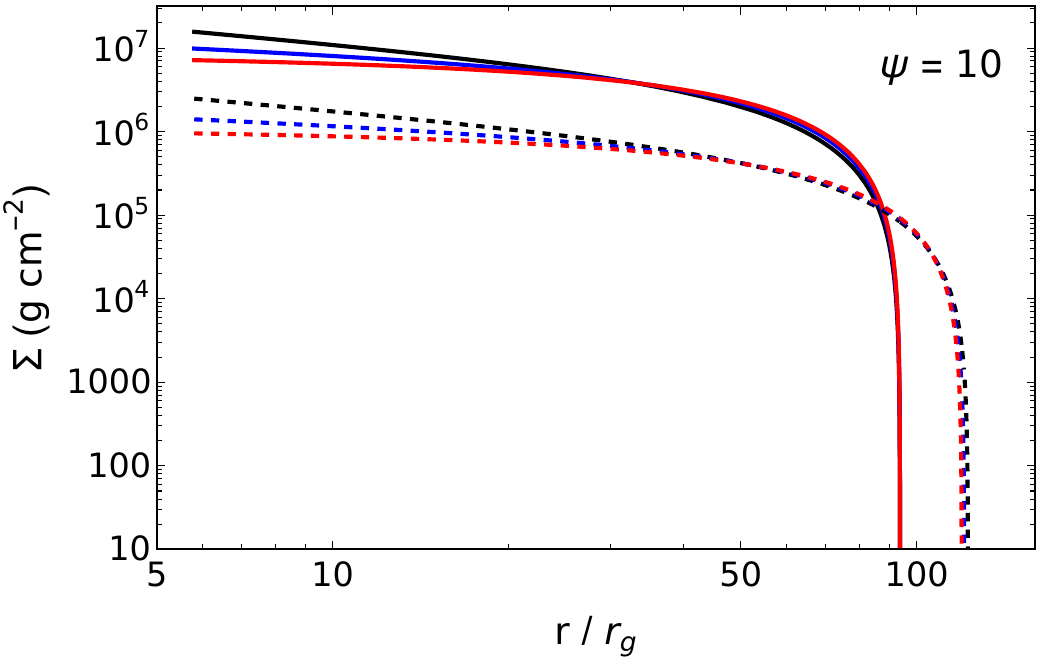}}
\caption{
Radial profile of the surface densities of the $(\beta_+,~\omega_+)$ solution at two different times for the three different $\epsilon_{\rm rad}$ and the two different $\psi$. Panels (a) and (b) correspond to the two cases of $\psi = 0$ and $\psi=10$, respectively. In all panels, the different colors represent three different values of $\epsilon_{\rm rad} = 0.1, 0.5,$ and 1, while the solid and dashed lines indicate the normalized times: $\tau = 1$ and 5, respectively.
}
\label{fig:sig}
\end{figure*}

Figure \ref{fig:sig} shows the radial profile of the surface density at different times for the $(\beta_+,~\omega_+)$ solution, considering three values of $\epsilon_{\rm rad} =$ 0.1, 0.5, and 1, and two values of $\psi =$ 0 and 10. In all cases, the outer radius expands by viscous spreading \cite{1974MNRAS.168..603L}.
The figure also shows that the surface density in the inner region of the disk decreases with time due to the presence of wind, while the magnetic brake suppresses the outward radial flow driven by viscosity.

%
%

%
\subsection{Mass Accretion and Loss Rates}
%
The mass accretion rate is given by using equations~(\ref{eq:azphi}), (\ref{eq:cs6}), and (\ref{eq:rsvr}) as
\begin{eqnarray}
\dot{M}
&=&
-2\pi {r} \Sigma v_r = 
4\pi \bar{\alpha}_{r\phi}
\frac{r^{-(1+\psi/2)}}{\Omega} 
\frac{\partial}{\partial r}
\biggr[
r^{2+\psi/2} c_s^2 \Sigma
\biggr].
\label{eq:mdacc}
\end{eqnarray}
Using the self-similar solution (\ref{eq:sdsol}), we rewrite equation~(\ref{eq:mdacc}) as 
\begin{eqnarray}
\dot{M}
&=&
3\pi
\left[\frac{4}{15}\omega\left(1+\frac{1}{5}\left[\psi + 7-\frac{\beta}{\omega}\right]\right)^{-1}\right]^{5/2}
\frac{\Sigma_0{r_0^2}}{t_0}
\left(
\frac{	r}{r_0} 
\right)^{-(\psi+5-\beta/\omega)/2}
\nonumber \\
&\times&
\biggr[\left(\frac{\beta}{\omega}-4\right)-(8+\psi)\xi^{
	2+\frac{2}{5}
	\left(
	\psi + 7 - \frac{\beta}{\omega}
	\right)}
\biggr] 
\left[ 
1 
- 
\xi^{2+\frac{2}{5}
\left(
\psi + 7 - \frac{\beta}{\omega}
\right)} 
\right]^{3/2}
\tau^{\omega(\psi+7)-8\beta/3}.
\label{eq:mdacc2}
\end{eqnarray}
Now we are interested in the mass accretion rate at the inner boundary radius near black hole, giving us the condition $\xi\ll1$ because $r_{\rm in}/r_{\rm out}\ll1$. The power law index of $\xi$ in equation~(\ref{eq:mdacc2}), $2+\frac{2}{5}\left(\psi + 7 - \frac{\beta}{\omega}\right)$, is always positive for any values of $\epsilon_{\rm rad}$ and $\psi$. This is because $2+\frac{2}{5}\left(\psi + 7 - \frac{\beta}{\omega}\right)= 3 / (5~ \omega_{-}) > 0$, where we have used equations~(\ref{eq:omega}) and (\ref{eq:beta}).
Equation~(\ref{eq:mdacc2}) at $\xi\ll1$ is then reduced to 
\begin{eqnarray}
\dot{M}
&=&
3\pi
\left[\frac{4}{15}\omega\left(1+\frac{1}{5}\left[\psi + 7-\frac{\beta}{\omega}\right]\right)^{-1}\right]^{5/2}
\frac{\Sigma_0{r_0^2}}{t_0}
\left(
\frac{	r}{r_0} 
\right)^{-(\psi+5-\beta/\omega)/2}
\biggr[
\frac{\beta}{\omega}
-4
\biggr] \nonumber \\
\nonumber \\
&& \times \,
\tau^{\omega(\psi+7)-8\beta/3},
\label{eq:mdacc3}
\end{eqnarray}
where $\beta/\omega-4$ = $[1+\psi+\sqrt{(1+\psi)^2+8(1-\epsilon_{\rm  rad})\left(3+\psi\right)}]/2$ for $\omega=\omega_+$ is always positive for any values of $\epsilon_{\rm rad}$ and $\psi$. Assuming that $\dot{M}\propto\tau^{n_{\rm acc}}$, we estimate the power law index of time as 
\begin{equation}
n_{\rm acc} = \frac{3(7+\psi)\omega-8\beta}{3} = 
-\frac{15+\psi+4\sqrt{(1+\psi)^2+8(1-\epsilon_{\rm  rad})\left(3+\psi\right)}}{15+\psi+\sqrt{(1+\psi)^2+8(1-\epsilon_{\rm  rad})\left(3+\psi\right)}}.
\label{eq:nacc}
\end{equation}
We confirm that $n_{\rm acc}=-19/16$ under the condition that there is no wind ($\epsilon_{\rm rad}=1$, $\psi=0$), consistent with the result of Cannizzo et al. (1990) \cite{1990ApJ...351...38C}.
In contrast, $n_{\rm acc} = -5/2$ in the limit $\psi \rightarrow \infty$, regardless of $\epsilon_{\rm rad}$.

%
%

Next, we derive the mass loss rate from the accretion disk. Using equations (\ref{eq:sigdotw}), (\ref{eq:ssc1}), and (\ref{eq:sdsol}), we obtain the mass loss rate as
\begin{eqnarray}
\dot{M}_{\rm w}
 &=& 
 \int_{r_{\rm in}}^{r_{\rm out}} \dot{\Sigma}_{\rm w} 2\pi r \, dr  
 = 
 6\pi (1-\epsilon_{\rm rad})
 \left(1+\frac{\psi}{3}\right)
\Lambda
 \int_{r_{\rm in}}^{r_{\rm out}} \Sigma^{5/3} \, dr 
 \label{eq:Mw_radial}\\
\nonumber \\ 
&=& 
18 \pi 
\left(1 - \epsilon_{\rm rad}\right)
\left(1+\frac{\psi}{3}\right)
\left[\frac{4}{15}\omega\left(1+\frac{1}{5}\left[\psi + 7-\frac{\beta}{\omega}\right]\right)^{-1}\right]^{5/2}
\frac{\Sigma_0 r_0^2}{t_0} 
\,  \nonumber \\
\nonumber \\
&& \times \, \tau^{2\omega-\frac{5}{3}\beta}
\,
G_3(\xi_{\rm in}),
\label{eq:mdw}
\end{eqnarray}
where 
\begin{eqnarray}\label{eq:G3t}
G_3(\xi_{\rm in})
\equiv
\int_{\xi_{\rm in}(\tau)}^1 \xi^{-\left( \psi+6-\frac{\beta}{\omega} \right)}  
\left[1-\xi^{2+ \frac{2}{5}\left(\psi+7-\frac{\beta}{\omega}\right)}\right]^{5/2} \, d\xi
\end{eqnarray}
is a dimensionless function of $\xi_{\rm in}(\tau)$. Similar to equation (\ref{eq:G1split}), we divide the domain of integration into two parts: $\{\xi_{\rm in}(\tau),~\xi_{\rm tr}\}$ and $\{\xi_{\rm tr},~1\}$. Equation (\ref{eq:G3t}) is then written as
\begin{eqnarray}\label{eq:G3split}
	G_3(\xi_{\rm in}) = \frac{\xi_{\rm tr}^{-(\psi+5-\beta/\omega)}-\xi_{\rm in}^{-(\psi+5-\beta/\omega)}}{-(\psi+5-\beta/\omega)} + \int_{\xi_{\rm tr}}^1 \xi^{-\left( \psi+6-\frac{\beta}{\omega} \right)}  
	\left[1-\xi^{2+ \frac{2}{5}\left(\psi+7-\frac{\beta}{\omega}\right)}\right]^{5/2} \, d\xi.
\end{eqnarray}
The second term on the right hand side is finite. For $\omega = \omega_{+}$ case, the power index $-(\psi+5-\beta/\omega) = \left(\sqrt{(1+\psi)^2+8(1-\epsilon_{\rm  rad})\left(3+\psi\right)} - \psi -1 \right)/2$, which is positive for $\epsilon_{\rm rad} \leq 1$ and $\psi \geq 0$. This implies that, in the limit $\xi_{\rm in} \rightarrow 0$, the first term in equation~(\ref{eq:G3split}) approaches a finite value, and thus $G_3(\xi_{\rm in})$ remains bounded. Therefore, $G_3(\xi_{\rm in})$ approaches a steady, finite value at late times.

Panel (a) of Figure~\ref{fig:g3g4} shows the dependence of $G_3(\xi_{\rm in})$ on $\xi_{\rm in}$ for $\omega=\omega_{+}$, estimated using equation (\ref{eq:G3t}). 
From the panel, it is clear that $G_3(\xi_{\rm in})$ approximately independent of time for $\xi_{\rm in} < 0.1$, consistent with the behavior expected from equation~(\ref{eq:G3split}). As a result, the mass loss rate scales as $\tau^{2\omega-(5/3)\beta}$ when $G_3(\xi_{\rm in})$ is constant. Assuming that $\dot{M}_{\rm w} \propto \tau^{n_{\rm w}}$, we obtain the power law index of time as
\begin{equation}
n_{\rm w}
=
2\omega-\frac{5}{3}\beta = 
 -\frac{33+ 5 \psi +5\sqrt{(1+\psi)^2+8(1-\epsilon_{\rm  rad})\left(3+\psi\right)}}{2(15+\psi+\sqrt{(1+\psi)^2+8(1-\epsilon_{\rm  rad})\left(3+\psi\right)})},
\label{eq:nw}
\end{equation} 
where equations~(\ref{eq:ssp1}) and (\ref{eq:omega}) were used for the derivation. In the limit $\psi \rightarrow \infty$, $n_{\rm w} = -5/2$, which is the same as in the case of $n_{\rm acc}$.

%
\subsection{Radiative Flux Density and Bolometric Luminosity}
%
Substituting equations~(\ref{eq:azphi}) and (\ref{eq:sdsol}) into equation (\ref{eq:qvisf}) gives
\begin{eqnarray}
Q_{+}
&=&
\frac{3}{4}
\frac{GM\Sigma_0}{r_0 t_0}
(\psi+3)
\left[\frac{4}{15}\omega\left(1+\frac{1}{5}\left[\psi + 7-\frac{\beta}{\omega}\right]\right)^{-1}\right]^{5/2}
\tau^{-5\beta/3-4\omega} \left[ 1 - \xi^{2 + \frac{2}{5} \left( \psi + 7 -\frac{\beta}{\omega}\right)} \right]^{5/2} \nonumber \\
&& \times \,
\xi^{-(\psi+11-\beta/\omega)}.
\label{eq:qplus}
\end{eqnarray}
According to the Stefan-Boltzmann law, the radiative flux emitted from the disk surface can be written as $\mathcal{F}=2\sigma T_{\rm eff}^4$, where $T_{\rm eff}$ is the surface temperature of the disk. Equating $\mathcal{F}$ with the radiative cooling flux in equation (\ref{eq:qrad}) gives the relation between the surface temperature of the disk and the disk mid-plane temperature $T_{\rm eff}=(Q_{\rm rad}/[2\sigma])^{1/4}=(32/[3\kappa_{\rm es}\Sigma])^{1/4} T$, and alternatively yields $T_{\rm eff}$ as the following equation:
\begin{eqnarray}
T_{\rm eff}(r,t) 
=
T_{\rm eff,0}\,
\tau^{-(5\beta/12+\omega)}
\left[ 1 - \xi^{2 + \frac{2}{5} \left( \psi + 7 -\frac{\beta}{\omega}\right)} \right]^{5/8}
\xi^{-(\psi+11-\beta/\omega)/4},
\label{eq:teff}
\end{eqnarray}
where  
\begin{eqnarray}\label{eq:Teff0}
T_{\rm eff,0}
\equiv
\left(
\frac{3}{8}
\right)^{1/4}
\left(
\frac{4}{15}
\right)^{5/8}
\left[
\epsilon_{\rm rad}^{2/5}(\psi+3)^{2/5}
\omega\left(1+\frac{1}{5}\left[\psi + 7-\frac{\beta}{\omega}\right]\right)^{-1}\right]^{5/8}
\left(
\frac{1}{\sigma}
\frac{GM\Sigma_0}{r_0 t_0}
\right)^{1/4}
\end{eqnarray}
and equations~(\ref{eq:qradf}) and (\ref{eq:qplus}) were used for the derivation.
The observed flux density is 
\begin{equation}
S_\nu=
4\pi 
\frac{\cos{i}}{D^2} 
 \frac{h\nu^3}{c^2} 
\int_{r_{\rm in}}^{r_{\rm out}}
\frac{1}{\exp\left[h\nu/k_{\rm B} T_{\rm eff}(r,t)\right]-1} r
\, dr,
\end{equation}
where $i$ is the inclination angle measured from the disk’s normal vector and $D$ is the luminosity distance (see \cite{1981ARA&A..19..137P,2002apa..book.....F,2008bhad.book.....K} for a review). We assume a face-on viewing geometry, corresponding to $i=0^{\circ}$ (i.e., $\cos{i} = 1$). The bolometric luminosity is then given by
\begin{eqnarray}
L_{\rm bol}
&=&
\int_0^\infty
L_\nu
\,
d\nu
=
16\pi^2 
 \frac{h}{c^2} 
\,
\int_{r_{\rm in}}^{r_{\rm out}}
\,
\left[
\int_0^\infty
\frac{\nu^3}{\exp\left[h\nu/k_{\rm B} T_{\rm eff}(r,t)\right]-1} 
d\nu
\right]
r\, dr
\nonumber \\
&=&
\frac{32}{15}\pi^6 
\frac{h r_0^2}{c^2}
\left(\frac{k_{\rm B}T_{\rm eff,0}}{h}\right)^4
\tau^{-5\beta/3} 
\,
G_4(\xi_{\rm in}), 
\label{eq:lumt}
\end{eqnarray}
where $L_\nu = 4 \pi D^2 S_{\nu} $ is the spectral luminosity,
\begin{eqnarray}\label{eq:G4t}
G_4(\xi_{\rm in})
\equiv
\int_{\xi_{\rm in}}^{1}
\left[ 1 - \xi^{2 + \frac{2}{5} \left( \psi + 7 -\frac{\beta}{\omega}\right)} \right]^{5/2}
\xi^{-(\psi+8-\beta/\omega)}
\,d\xi
\end{eqnarray}
is a dimensionless function of $\xi_{\rm in}(\tau)$, and we used the integral formula as $\int_0^\infty\,x^3/(e^x-1)\, dx=\pi^4/15$ for the derivation. 
Similar procedure to equation~(\ref{eq:G1split}) allows us to divide the integration domain in equation~(\ref{eq:G4t}) into two parts through $\xi_{\rm tr}$.
Consequently, equation~(\ref{eq:G4t}) becomes
\begin{equation}
\label{eq:G4split}
G_4(\xi_{\rm in})
=
\frac{\xi_{\rm in}^{-(\psi+7-\beta/\omega)}}{(\psi+7-\beta/\omega)}
-
\frac{\xi_{\rm tr}^{-(\psi+7-\beta/\omega)}}{(\psi+7-\beta/\omega)} +
\int_{\xi_{\rm tr}}^1 \xi^{-\left( \psi+8-\frac{\beta}{\omega} \right)}
\left[1-\xi^{2+ \frac{2}{5}\left(\psi+7-\frac{\beta}{\omega}\right)}\right]^{5/2} 
d\xi,
\end{equation}
where the first term (i.e., inner-edge term) is the only part that depends on time through $\xi_{\rm in} \propto \tau^{-\omega}$, while the remaining terms (i.e., the outer-edge term) are constant and time-independent. This decomposition will be useful in later paragraphs.
For the case $\omega = \omega_{+}$, the exponent $(\psi + 7 - \beta/\omega) = (\psi + 5 - \sqrt{(1 + \psi)^2 + 8(1 - \epsilon_{\rm rad})(3 + \psi)})/2$ is positive. This indicates that $G_4(\xi_{\rm in})$ continues to grow as $\xi_{\rm in}^{-(\psi + 7 - \beta/\omega)}$ in the limit $\xi_{\rm in} \rightarrow 0$. This implies that the luminosity diverges as $r_{\rm in} \rightarrow 0$, so a finite inner truncation is required to obtain a physically reasonable luminosity.

%
\begin{figure*}
	\centering
	\subfigure[]{\includegraphics[scale = 0.6]{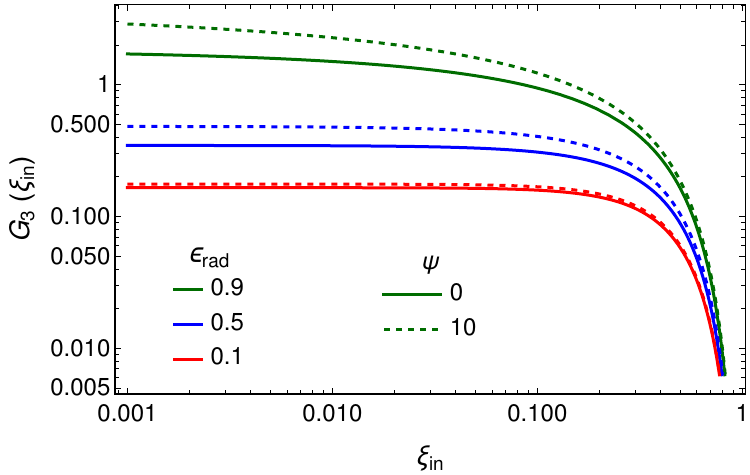}}
	\subfigure[]{\includegraphics[scale = 0.6]{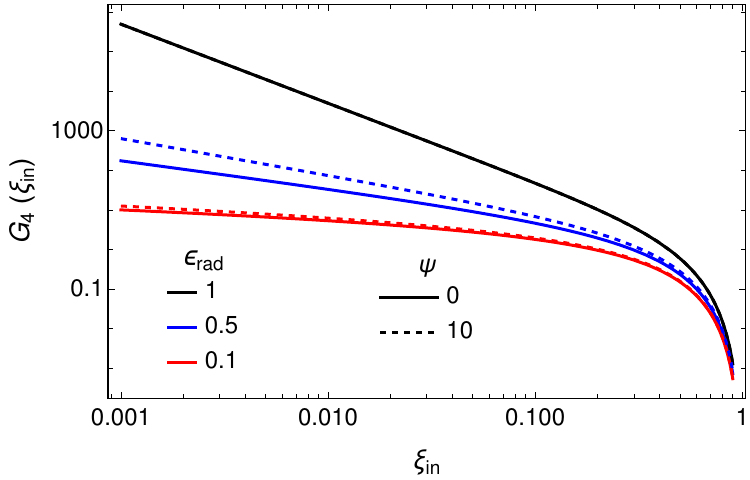}}
	\caption{
		Dependence of $G_3(\xi_{\rm in})$ and $G_4(\xi_{\rm in})$ on $\xi_{\rm in}$ for the case of $\omega=\omega_+$. Panels (a) and (b) shows $G_3$ and $G_4$, respectively. For all the panels, the different color represents three different $\epsilon_{\rm rad}$ and the solid and dashed lines show the $\psi=0$ and $\psi=10$ cases, respectively.
	}
	\label{fig:g3g4}
\end{figure*}
%

Panel (b) of Figure~\ref{fig:g3g4} shows that $G_4(\xi_{\rm in})$ is not constant over the entire range of $\xi_{\rm in}$; that is, $G_4(\xi_{\rm in})$ is time-dependent. This behavior is also expected from equation~(\ref{eq:G4split}). Assuming that $L_{\rm bol}\propto \tau^{n_{\rm l}}$, we get
\begin{eqnarray}
\label{eq:nlt}
n_{\rm l} = \frac{\partial \ln{G_4(\xi_{\rm in}(\tau))}}{\partial \ln{\tau}}-\frac{5}{3}\beta.
\end{eqnarray}
According to equation~(\ref{eq:G4split}), $G_4(\xi_{\rm in}) \propto \xi_{\rm in}^{-(\psi + 7 - \beta/\omega)}$ in the limit $\xi_{\rm in} \rightarrow 0$.
Substituting this into equation~(\ref{eq:nlt}), and taking into account that $\xi_{\rm in} = \sqrt{r_{\rm in}/r_0} \, \tau^{-\omega}$, gives
\begin{equation}
n_{\rm l} 
= -\frac{8}{3}\beta + \omega (\psi + 7) 
=-\frac{15 + \psi + 4\sqrt{(1 + \psi)^2 + 8(1 - \epsilon_{\rm rad})(3 + \psi)}}{15 + \psi + \sqrt{(1 + \psi)^2 + 8(1 - \epsilon_{\rm rad})(3 + \psi)}}
\label{eq:nl}
\end{equation}
in the limit $\tau\to\infty$, where equations~(\ref{eq:omega}) and (\ref{eq:beta}) were used for the derivation. This is identical to the power-law index of the mass accretion rate, $n_{\rm acc}$.

\begin{figure}
	\centering
	\subfigure[]{\includegraphics[scale = 0.6]{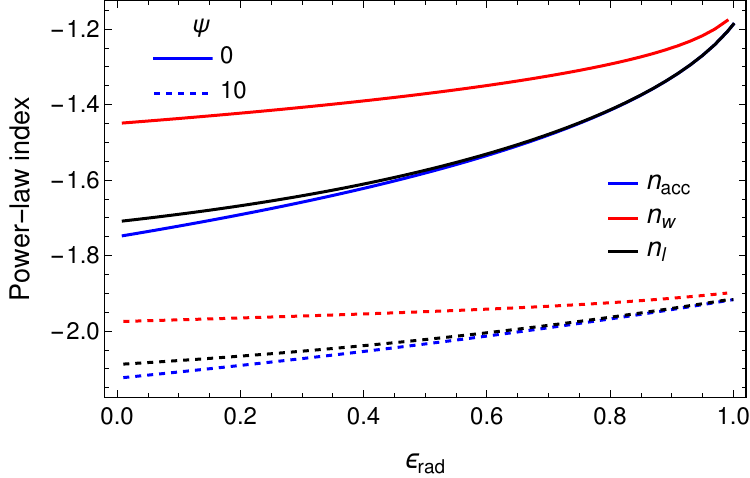}}
	\subfigure[]{\includegraphics[scale = 0.6]{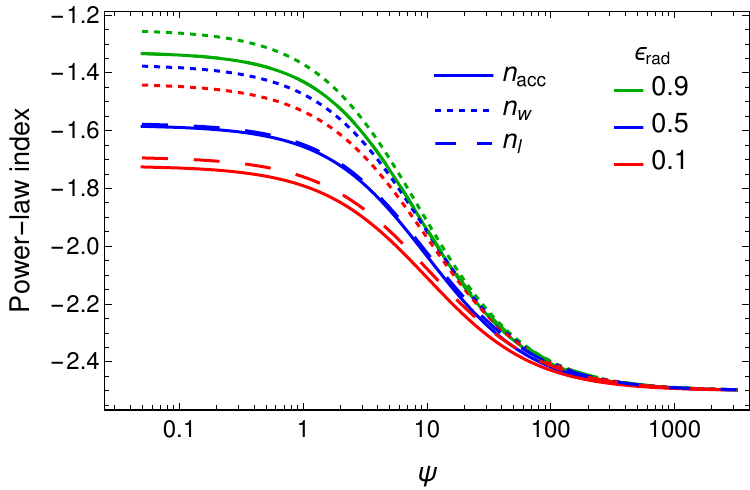}}
	\caption{\label{fig:slope} 
		The power law indices of the time evolution of mass accretion rate ($n_{\rm acc}$), wind mass loss rate ($n_{\rm w}$), and the bolometric luminosity ($n_{\rm l}$, estimated using equation \ref{eq:nlt} at late times) is shown. Panel (a) shows the power indices for various $\epsilon_{\rm rad}$ with two values of $\psi = 0$ and $10$. Panel (b) shows the power indices for various $\psi$ with three values of $\epsilon_{\rm rad} = 0.1,~ 0.5$, and $0.9$. 
	}
\end{figure}

Panel (a) of Figure \ref{fig:slope} illustrates the dependence of three power-law indices, $n_{\rm acc}$, $n_{\rm w}$, and $n_{\rm l}$, on $ \epsilon_{\rm rad}$ for different values of $\psi$. It can be observed that $n_{\rm w}$ is the highest among the three indices, indicating that the mass loss rates are less steep compared to the other two rates in the range $0 < \epsilon_{\rm rad} < 1$. Panel (b) of Figure \ref{fig:slope} depicts the dependence of $\psi$ on the three power-law indices for different values of $\epsilon_{\rm rad}$. From this panel, it is evident that $n_{\rm acc}$ and $n_{\rm l}$ have steeper (more negative) slopes than $n_{\rm w}$ for a given $\psi$. Additionally, all three indices asymptote to $-5/2$ as $\psi \rightarrow \infty$.

In the canonical self-similar solution of Cannizzo et al. (1990) \citep{1990ApJ...351...38C} with a vanishing torque at the inner boundary, the bolometric luminosity scales with the mass-accretion rate. Therefore, the temporal power-law index of $L_{\rm bol}$ is equal to that of $\dot{M}$. Our self-similar solutions maintain this relation when $\xi_{\rm in}$ goes to zero. In this limit, the inner-edge term of $G_4(\xi_{\rm in})$ dominates, and thus  the power-law indices of luminosity and accretion rate become equal.

However, since the inner radius is finite and time-dependence of $G_4$ depends on how fast $\xi_{\rm in}$ shrinks towards zero, the two power-law indices can differ. For $\epsilon_{\rm rad}\ll1$, much of the heating energy powers the wind. The temperature and surface-density gradients remain shallow, $\xi_{\rm in}$ shrinks slowly, and both terms of $G_4$ tend to be comparable.  A similar effect occurs when angular momentum is not effectively removed due to weaker magnetic braking. Consequently, when the condition of $\epsilon_{\rm rad} \ll 1$ and $0\le\psi \lesssim 1$ is satisfied simultaneously, the outer-edge term of $G_4$ becomes comparable to the inner-edge term. This causes a clear deviation between $n_{\rm l}$ and $n_{\rm acc}$, as shown in Figure 5(b).

In the broader parameter space where $0.1 \lesssim \epsilon_{\rm rad} \le 1$ or magnetic braking is strong, the inner-edge term dominates $G_4$, causing $n_{\rm l}$ to approach $n_{\rm acc}$. Note that both indices always converge at the limit of $\tau\to\infty$ as shown in equation~(\ref{eq:nl}), regardless of the values of $\epsilon_{\rm rad}$ and $\psi$. Therefore, a significant difference between $n_{\rm l}$ and $n_{\rm acc}$ only appears during the early to middle stages of disk evolution when $\epsilon_{\rm rad} \ll 1$ and $0\le \psi \lesssim 1$. 

%
%

\begin{figure}
	\centering
		\subfigure[]{\includegraphics[scale = 0.6]{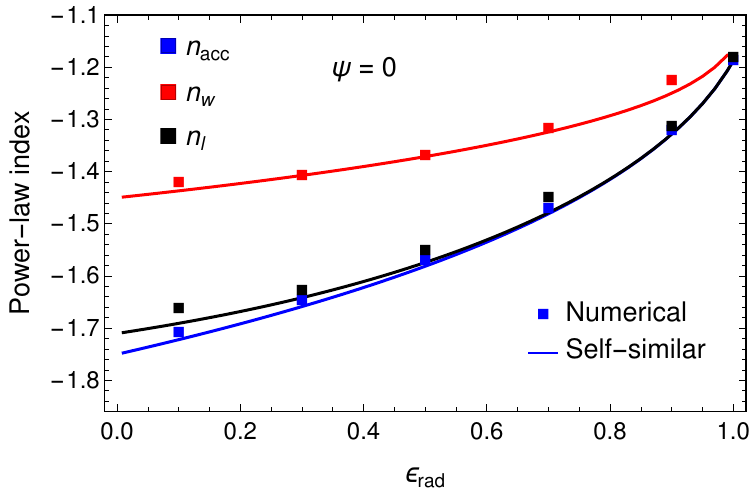}}
		\subfigure[]{\includegraphics[scale = 0.6]{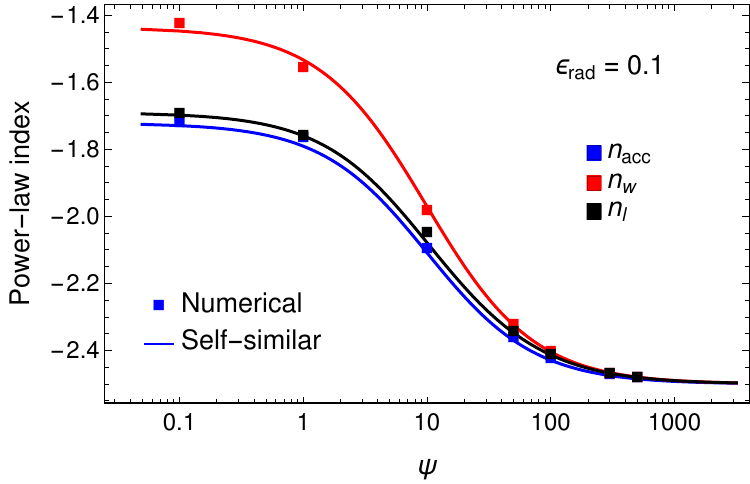}}\\
		\subfigure[]{\includegraphics[scale = 0.6]{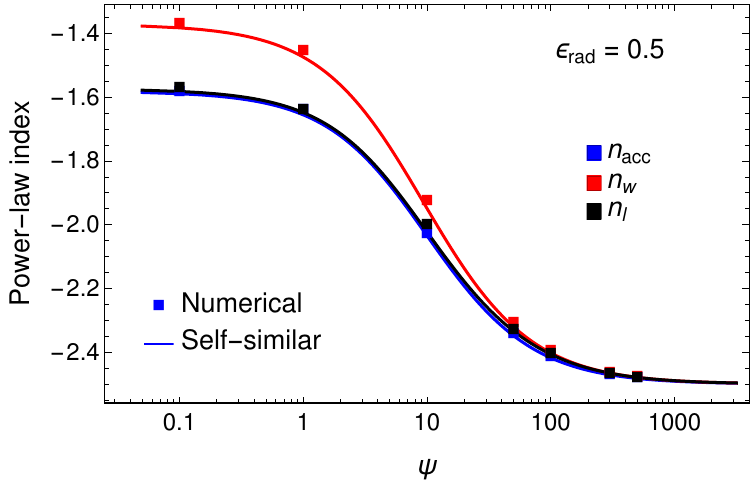}}
		\subfigure[]{\includegraphics[scale = 0.6]{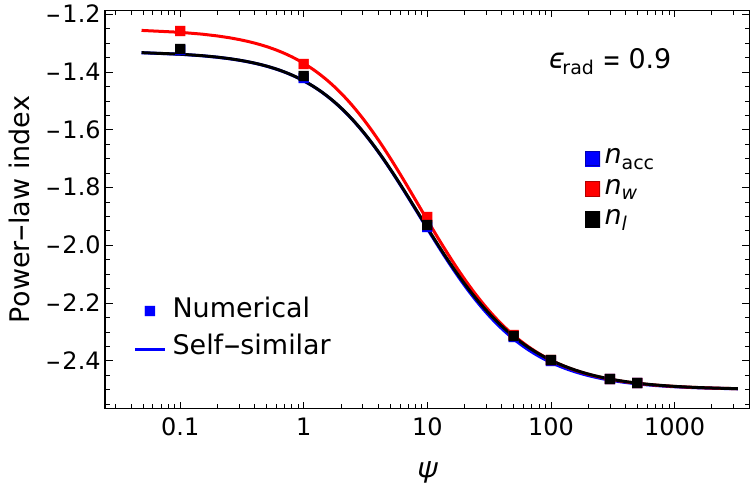}}
	\caption{\label{fig:slope_comp} 
		Comparison of $n_{\rm acc}$, $n_{\rm w}$ and $n_{\rm l}$ obtained from numerical simulations at $t/\tau_0 = 150$ with the self-similar solution. Panel (a) shows the dependence on $\epsilon_{\rm rad}$ with $\psi = 0$, whereas panels (b), (c), and (d) show the dependence on $\psi$ with $\epsilon_{\rm rad} = $ 0.1, 0.5, and 0.9, respectively. 
        The time normalization parameter $\tau_0 = 14.3$ yr corresponds to the viscous timescale for a TDE disk with scale height of 0.01 at the circularization radius, formed from the tidal disruption of a solar mass star by a black hole of mass $10^6 M_{\odot}$ (see equation 19 in Tamilan et al. 2024 \cite{2024ApJ...975...94T}). 
	}
\end{figure}

%
\subsection{Comparison between self-similar and numerical solutions}
%

Figure \ref{fig:slope_comp} shows a comparison of $n_{\rm acc}$, $n_{\rm w}$, and $n_{\rm l}$ obtained from numerical simulations at late times with the corresponding self-similar solutions. Panel (a) illustrates their dependence on $\epsilon_{\rm rad}$ with $\psi = 0$, while the remaining three panels depict their dependence on $\psi$ for different values of $\epsilon_{\rm rad}$. In all panels, the numerically estimated power-law indices show good agreement with the self-similar solutions. Our self-similar and numerical solutions also show $n_{\rm l}\simeq n_{\rm acc}$, except for the regime of ${\epsilon}_{\rm rad}\ll1$ and $0\le\psi\lesssim1$.

The mass accretion rates and bolometric light curves decrease more rapidly compared to the mass loss rates. Additionally, it is observed that both the self-similar and numerical solutions converge to $-5/2$ for large values of $\psi$.

%
\subsection{Integrated Optical-UV Luminosity and Its Instantaneous slopes}
\label{sec:OUV}
%

In this section, we compute the time evolution of the integrated disk luminosity within the optical–ultraviolet (UV) frequency band, corresponding to typical observations of TDEs. The frequency range adopted here, from $4.61 \times 10^{14}$ to $1.76 \times 10^{15}$ Hz, is based on the wavelength coverage of the Swift UVOT filters (170–650 nm) \cite{uvotweb}. The optical-UV luminosity is given by
\begin{equation}
L_{\mathrm{OUV}}(t) =
16\pi^2
\frac{h}{c^2}
\int_{r_{\mathrm{in}}}^{r_{\mathrm{out}}(t)}
\left[ \int_{\nu_{\min}}^{\nu_{\max}}
\frac{\nu^3}{\exp\left[h\nu/k_{\rm B} T_{\rm eff}(r,t)\right]-1} 
d\nu \right] r\, dr,
\label{eq:LOUV}
\end{equation}
where $T_{\mathrm{eff}}(r,t)$ is the disk surface temperature determined from equation~(\ref{eq:teff}), the outer radius evolves as $r_{\mathrm{out}}(t) = r_0\,\tau^{2\omega}$ with $\tau = t / t_0$, and the lower and upper limits of integration, $\nu_{\min}$ and $\nu_{\max}$, are fixed by the filter bandpass.


The resulting light curves $L_{\mathrm{OUV}}(t)$ exhibit a time evolution that cannot be captured by a single power-law function. Instead, the decay slope varies gradually over time. This behavior originates from three key physical mechanisms: the outward expansion of the disk surface area, the time-dependent evolution of $T_{\rm eff}(r,t)$, and the shifting spectral regime of the emission. First, the increase of the disk's outer radius leads to a growing surface area, which enhances the total luminosity during the early stages even if the temperature declines. Second, unlike steady-state accretion disks where the effective temperature profile follows $T_{\rm eff} \propto \dot{M}^{1/4} r^{-3/4}$ \citep{LodatoRossi2011}, our self-similar solution exhibits a coupled dependence of $T_{\rm eff}$ on both radius and time, driven by the combined effects of viscous diffusion and mass loss. 
As a consequence, the radius $r_{\mathrm{obs}}(t)$ at which the local temperature equals a fixed observational value $T_{\mathrm{obs}}$, typically $\sim 2 \times 10^4$ K (e.g., \cite{Holoien2016,vanVelzen2020}), evolves nonlinearly in time. This radius determines the dominant contribution to the emission in the optical-UV band. Third, while the observational frequency range is fixed, whether the emitting gas is in the Rayleigh–Jeans or Wien regime depends on the self-similar surface temperature. When $kT_{\mathrm{eff}} \gg h\nu$, the emission is in the Rayleigh–Jeans limit and scales linearly with temperature. When $kT_{\mathrm{eff}} \ll h\nu$, the emission is suppressed exponentially, characteristic of the Wien regime.

 \begin{figure*}
  \centering
\subfigure[]{\includegraphics[scale = 0.57]{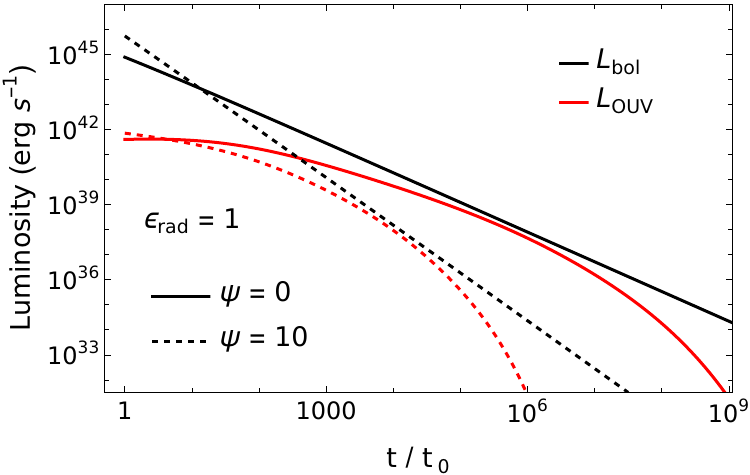}}
\subfigure[]{\includegraphics[scale = 0.57]{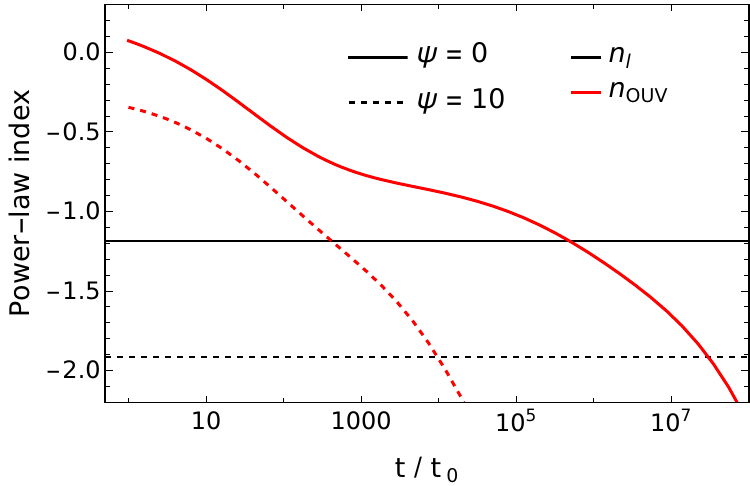}}
\\
\subfigure[]{\includegraphics[scale = 0.57]{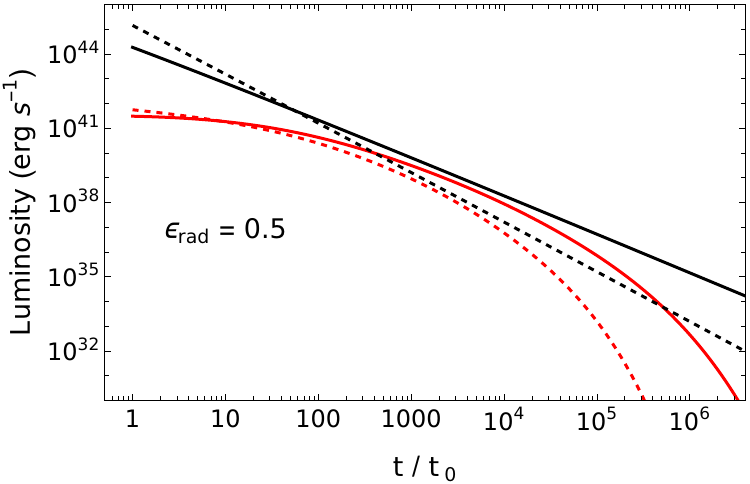}}
\subfigure[]{\includegraphics[scale = 0.57]{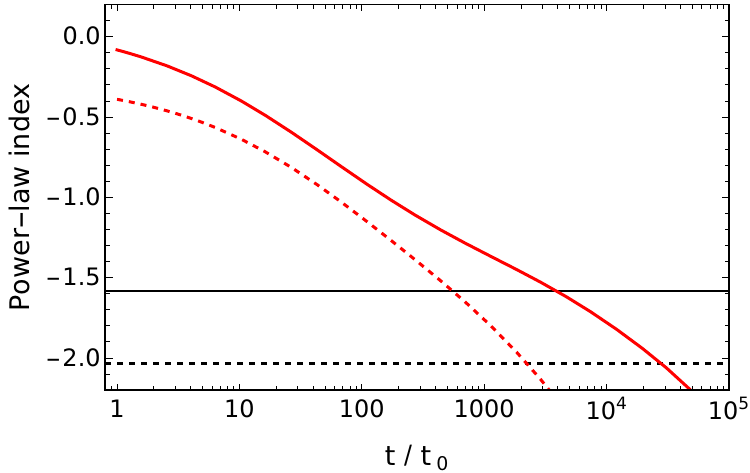}}
\\
\subfigure[]{\includegraphics[scale=0.57]{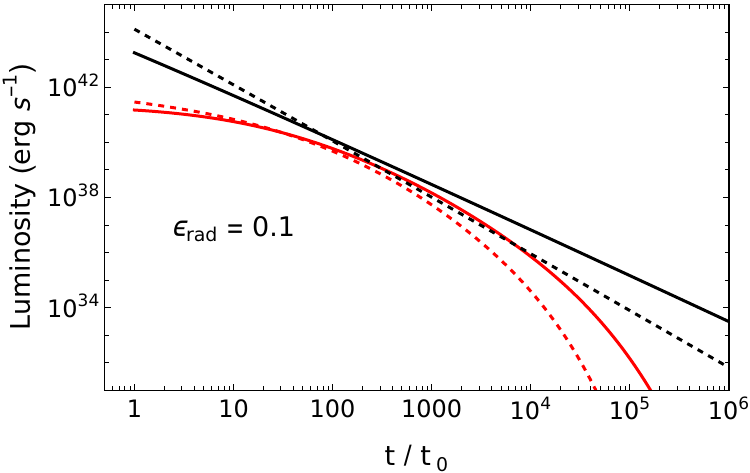}}
\subfigure[]{\includegraphics[scale = 0.57]{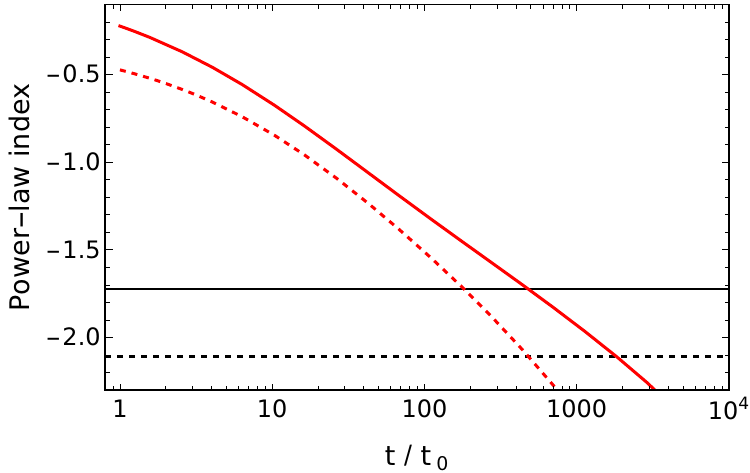}}
  \caption{
Time evolution of the integrated optical–UV luminosity $L_{\mathrm{OUV}}(t)$ for the wavelength band 170–650 nm. The left column (panels (a), (c), (e)) presents $L_{\mathrm{OUV}}(t)$ and the bolometric luminosity, $L_{\rm bol}$, for three flux ratios: $\epsilon_{\rm rad}=1$ (top panels), $0.5$ (middle), and $0.1$ (bottom). Black curves denote the bolometric output, red curves the optical–UV band. Solid curves are computed without a wind ($\psi=0$); dashed curves include a strong wind ($\psi=10$). The right column (panels (b), (d), (f)) shows the instantaneous power-law index $n_{\mathrm{OUV}}$ obtained by numerically differentiating the corresponding curves in the left column. The horizontal black solid and dashed lines indicate the power-law index of the bolometric luminosity for each parameter set.
}
\label{fig:OUVcurves}
\end{figure*}

The left panels of Figure~\ref{fig:OUVcurves} display the computed light curves for six representative models, spanning cooling-to-heating flux ratios $\epsilon_{\rm rad} = 1$, 0.5, and 0.1, and $\psi = 0$ and 10, where $\psi$ controls the strength of magnetic braking. All cases exhibit the three-phase evolution described above. The early-time luminosity is generally higher in models with larger $\psi$ because magnetic braking promotes mass accretion, which in turn enhances the release of gravitational energy through accretion. This leads to a higher midplane heating rate and a correspondingly elevated effective temperature at the disk surface. Even when $\epsilon_{\rm rad} = 1$, where no wind is launched, the early-time luminosity remains high because the initial disk mass is large even for $\psi = 0$, and the viscous heating rate increases with $\psi$. When $\epsilon_{\rm rad} < 1$, part of the thermal energy drives a wind instead of being fully radiated, but the enhanced heating still results in high surface temperatures and thus high luminosity at early times. As the system evolves, wind-driven mass loss can rapidly reduce the surface density, especially for models with low $\epsilon_{\rm rad}$ and high $\psi$, leading to a steep decline in luminosity. In contrast, no-wind models retain more disk mass and exhibit slower luminosity decay. Because the surface temperature profile and dominant emission radius evolve in a model-dependent and nontrivial way, $L_{\mathrm{OUV}}(t)$ cannot be described by a single power-law index. This complexity is important to consider when interpreting observed optical/UV TDE light curves, which often show a slowly varying decay rather than a single temporal index.

To quantify the time-varying decay rate, we introduce the instantaneous power-law index
\begin{equation}
n_{\rm{OUV}}(t) 
=
\frac{d\ln L_{\rm OUV}(t)}{d\ln t}
\end{equation}
and evaluate it numerically for the six representative models with $\epsilon_{\rm rad} = 1$, $0.5$, $0.1$ and $\psi = 0$, $10$. The results are shown in the right-hand panels of Figure~\ref{fig:OUVcurves}. All cases exhibit a gradual decrease in $n_{\mathrm{OUV}}(t)$ from near zero at early times, reflecting Rayleigh–Jeans–dominated emission with an expanding surface area, to values approaching $-2$ at late times as the emission transitions into the Wien regime. The bolometric luminosity power-law index $n_{\rm l}$, shown by horizontal black lines, remains constant and steeper throughout. Wind-dominated models display a more rapid rise in slope due to mass loss from the disk, whereas the higher cooling-to-heating flux ratio leads to slower power-law index evolution. 
The figure does not include observational data points; comparisons with observed light curve power-law indices of non-jetted TDEs are presented separately in Section~\ref{sec:discussion}.

%
\section{Discussion}
\label{sec:discussion}
%
 
There are three key parameters to characterize the evolution of a one-dimensional accretion disk with magnetically driven wind: ($\bar{\alpha}_{r\phi}$, $\bar{\alpha}_{z\phi}$, $\epsilon_{\rm rad}$), where note that $\psi$ can be an alternative to the vertical stress parameter, $\bar{\alpha}_{z\phi}$. The $\bar{\alpha}_{r\phi}$ and $\bar{\alpha}_{z\phi}$ have been introduced by using the $\alpha$-parameter prescription for MHD turbulence in radial and vertical directions.
The parameter $\epsilon_{\rm rad}$, representing the cooling-to-heating flux ratio, is introduced to close the basic equations and to control the vertical mass flux. 
Tamilan et al. (2024) \cite{2024ApJ...975...94T} performed numerical simulations of the basic equations in the context of TDE with $\bar{\alpha}_{z\phi} =0$ and $0.001$. They found that the power-law indices saturates at late times for $\bar{\alpha}_{z\phi} =0$ and gradually decreases with time without reaching a saturation value for $\bar{\alpha}_{z\phi} =0.001$. In contrast, our self-similar solutions suggest that $\bar{\alpha}_{z\phi} $ is a function of radius and time. In fact, substituting equations (\ref{eq:cs6}), and (\ref{eq:blambda}) into equation (\ref{eq:azphi}) yields

\begin{eqnarray}
\bar{\alpha}_{z\phi} 
&=& 
\Lambda^{1/2}
\psi 
\bar{\alpha}_{r\phi}^{1/2} 
(G M)^{-1/4} 
r^{1/4}\Sigma^{1/3} 
\nonumber \\
&=&
\Lambda^{1/2}
\psi 
\bar{\alpha}_{r\phi}^{1/2} 
(G M)^{-1/4} 
r_0^{1/4} \Sigma_0^{1/3}
 \left[\frac{4\omega}{15}\left(1+\frac{1}{5}\left(\psi+7-\frac{\beta}{\omega}\right)\right)^{-1}\right]^{1/2}  
\nonumber \\
&\times&
 \left(\frac{t}{t_0}\right)^{\omega(\psi+7-(8/3)\beta/\omega)/5} 
 \left(\frac{r}{r_0}\right)^{-\left(2\psi+9-2\beta/\omega\right)/20} 
 \nonumber \\
&\times&
\left[1-\left(\frac{r}{r_0}\right)^{(\psi+12-\beta/\omega)/5}\left(\frac{t}{t_0}\right)^{-2\omega(\psi+12-\beta/\omega)/5}\right]^{1/2},
\label{eq:azv}
\end{eqnarray}
where we adopted equations (\ref{eq:xi}) and (\ref{eq:sdsol}) for the second and third lines of the right hand side. The $\bar{\alpha}_{z\phi}$ decreases with time and increases with $\psi$ and $\epsilon_{\rm rad}$, though the increase with $\epsilon_{\rm rad}$ is modest. In numerical simulations by Tamilan et al. (2024) \cite{2024ApJ...975...94T}, the initial surface density was modeled as a Gaussian, peaking at the circularization radius, $r_{\rm cir} = 2 r_{\rm t}$, with a total disk mass equal to half the stellar mass. For a $10^6 M_{\odot}$ black hole and a solar mass star, this yields $r_{\rm cir} = 93.8 r_{\rm g}$ and a peak surface density of $\Sigma = 1.45 \times 10^{7}~{\rm g~cm^{-2}}$. Following this, we assume typical values of $r_0 = 100 r_{\rm g}$ and $\Sigma_0 = 10^{7}~{\rm g~cm^{-2}}$. For $\psi = 10$ and $\epsilon_{\rm rad} = 0.5$, equation (\ref{eq:azv}) results in
\begin{eqnarray}
\bar{\alpha}_{z\phi} 
&\sim&
1.5
\times 10^{-3} \left(\frac{\bar{\alpha}_{r\phi}}{0.1}\right)^{7/6} \left(\frac{M}{10^6 M_{\odot}}\right)^{-1/6} \left(\frac{r_0}{100 r_g}\right)^{1/4} 
\left(\frac{\Sigma_0}{10^7 ~{\rm g~cm^{-2}}}\right)^{1/3} 
\nonumber \\ 
&\times&
\left(\frac{r}{r_0}\right)^{1/20} \left[1-\left(\frac{r}{r_0}\right)^{7/5}\tau^{-7/30}\right]^{1/2}
\tau^{-23/60} 
.
\end{eqnarray}
Bai \& Stone (2013) \cite{2013ApJ...769...76B} performed MHD simulations of protoplanetary disks, finding maximum vertical stress $\alpha$-values ranging from $3\times 10^{-5}$ to $3 \times 10^{-3}$ for a plasma beta ranging between $10^{6}$ and $10^{3}$. Our estimated value for $\psi = 10$ falls within this range, consistent with results from numerical simulations. The vertical stress increases as plasma beta decreases, and similarly, $\bar{\alpha}_{z\phi}$ increases with $\psi$, implying that a higher $\psi$ corresponds to a stronger magnetic field in the disk.  Notably, as the simulation focuses on a protoplanetary disk, our study provides an alternative constraint on $\bar{\alpha}_{z\phi}$. A large $\psi$ corresponds to strong large-scale magnetic fields, which can induce a magnetocentrifugally driven wind. A smaller $\psi$, indicating lower vertical stress, suggests weaker poloidal magnetic fields, where MRI amplifies the magnetic field, driving the outflow via magnetic pressure with stochastic magnetic reconnections. It is important to note that there is no clear distinction between the magnetocentrifugal-driven and MRI-driven wind regimes.

Tabone et al. (2022) \cite{2022MNRAS.512.2290T} introduced the parameters $\alpha_{\rm SS}$ and $\alpha_{\rm DW}$ to represent the radial and vertical stresses, respectively, for their disk-wind model. A comparison with our $\alpha$ parameterization shows that $\bar{\alpha}_{r\phi}=(3/2)\alpha_{\rm SS}$ and $\bar{\alpha}_{z\phi}=(3/2)\alpha_{\rm DW}(H/r)$, leading to $\psi=\alpha_{\rm DW}/\alpha_{\rm SS}$, where equation~(\ref{eq:azphi}) was applied. They derived a self-similar solution by assuming that both the radial viscosity and the disk aspect ratio, $H/r$, are only a function of the disk radius, implying that the disk mid-plane temperature depends only on the radius. In contrast, in our model the disk mid-plane temperature has a more general form, i.e., it is a function of both the radius and the surface density, which is given by $T \propto r^{-1/2} \Sigma^{2/3}$. This difference causes the difference in the disk structure between their solution and ours. In the absence of wind, in the inner regions of the disk, their solution for $T \propto r^{-1/2}$ gives $\Sigma \propto t^{-3/2} r^{-1} $, while our solution for $(\beta_{+},~\omega_{+})$ gives $\Sigma \propto t^{-57/80} r^{-3/5} $. For the mass accretion rate, their solution gives that $\dot{M} \propto t^{-3/2} $, while our solution provides that $ \dot{M} \propto t^{-19/16} $ corresponding to the classical solution by Cannizzo et al. (1990) \cite{1990ApJ...351...38C}. In the presence of the wind, in both their model and ours, the mass accretion rate is steeper than in the no-wind case, and the mass loss rate is flatter than the mass accretion rate. However, the time evolution of the mass accretion and loss rates is different between their solutions and ours. For example, in a case where the magnetic braking works extremely efficiently (i.e., in the $\psi \rightarrow \infty $ case), they showed that the mass accretion and loss rates decrease exponentially with time, while our solution shows a $t^{-5/2} $ evolution for both rates.

As demonstrated above, our model exhibits distinct temporal evolutions of the mass accretion and loss rates compared to the model of Tabone et al. (2022) \cite{2022MNRAS.512.2290T}, regardless of the presence or absence of the wind. Notably, our model can also account for the power-law time indices observed in several TDE light curves. For instance, the UV/optical-derived bolometric light curve of AT2019qiz exhibits a rapid decay proportional to $t^{-2.54}$ \citep{2020MNRAS.499..482N}. This value was obtained by fitting blackbody spectra to interpolated multi-band photometry using SUPERBOL \citep{2018RNAAS...2..230N}, assuming spherical symmetry. Interestingly, our model predicts a similarly steep decline in the bolometric luminosity for strong magnetic braking (i.e., $\psi \gg 1$), where $n_{\rm l}$ approaches $-2.5$ (see panels (b)–(d) of Figure~\ref{fig:slope_comp}). This suggests that our model may qualitatively capture the rapid luminosity evolution seen in events such as AT2019qiz. We note, however, that the regime of $\psi \gg 1$ may correspond to a super-Eddington accretion flow (see also Tamilan et al. 2024 \cite{2024ApJ...975...94T}), which violates a fundamental assumption of our model. A more detailed investigation of this regime is beyond the scope of the present work and will be explored in a forthcoming study.

%
\subsection{Comparison with observed optical--UV TDE power-law indices} \label{sec:obs_comparison} 
%

\begin{figure}
\centering
\includegraphics[width=\columnwidth]{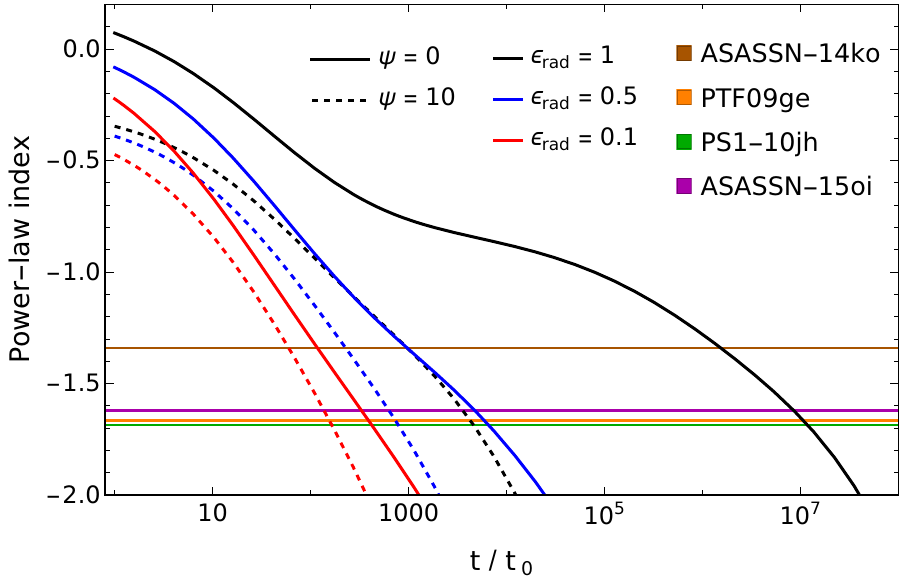}
\caption{
Instantaneous power-law index $n_{\mathrm{OUV}}(t)$ from the six theoretical models shown in Figure~\ref{fig:OUVcurves}, overplotted with horizontal lines representing observed optical--UV decay power-law indices for four non-jetted TDEs: PS1-10jh \cite{2012Natur.485..217G}, PTF09ge \cite{2014ApJ...793...38A}, ASASSN-15oi \cite{2016MNRAS.463.3813H}, and ASASSN-14ko \cite{2021ApJ...910..125P}.  
Solid and dashed lines indicate $\psi = 0$ and $\psi = 10$, respectively. The observational values cluster around $-1.6 \lesssim n_{\mathrm{OUV}} \lesssim -1.4$, which is consistent with the transitional phase predicted either by strong-wind models with a low cooling-to-heating ratio or by no-wind models with strong magnetic braking.
}
\label{fig:OUV_slope_obs}
\end{figure}

In Figure~\ref{fig:OUV_slope_obs}, we compare the predicted instantaneous power-law indices $n_{\mathrm{OUV}}(t)$ with observed decay power-law indices from several non-jetted TDEs. The figure overlays horizontal lines indicating the optical--UV power-law indices of PS1-10jh, PTF09ge, ASASSN-15oi, and ASASSN-14ko, derived from their light curves in the literature \cite{2012Natur.485..217G,2014ApJ...793...38A,2016MNRAS.463.3813H,2021ApJ...910..125P}.

The observational power-law indices fall within a relatively narrow range, $-1.6 \lesssim n_{\mathrm{OUV}} \lesssim -1.3$, with ASASSN-14ko at the shallow end ($1.33 \pm 0.03$) and others such as PS1-10jh and PTF09ge showing steeper decays consistent with the canonical $t^{-5/3}$ profile. ASASSN-15oi has been fitted by both exponential and power-law models, with the power-law index $t^{-1.62}$ yielding a slightly higher $\chi^2$ than the exponential fit \cite{2016MNRAS.463.3813H}. Here we adopt the power-law values for uniformity in comparison.

The colored model curves in Figure~\ref{fig:OUV_slope_obs} are identical to those in Figure~\ref{fig:OUVcurves}, showing the time-dependent power-law index evolution under various combinations of cooling-to-heating flux ratio $\epsilon_{\rm rad}$ and $\psi$, which characterizes the strength of magnetic braking. Notably, the model with $\epsilon_{\rm rad} = 1$ and $\psi = 10$, which corresponds to a fully cooled disk with strong magnetic braking but no wind, passes through the observed range of power-law indices at intermediate times $t/t_0 \sim 10^3$. In contrast, models with $\epsilon_{\rm rad} = 1$ and $\psi = 0$, that is, fully cooled disks without wind remain wind-free and generally produce shallower power-law indices at comparable times, reaching the observed range only at very late times. Meanwhile, models with $\epsilon_{\rm rad} < 1$ can launch disk winds regardless of $\psi$, and some of them also pass through the observational power-law index range, especially when $\psi$ is nonzero and reinforces angular momentum loss. This suggests that the optical–UV light curves of TDEs are broadly consistent with either thermally or magnetically driven winds, or with fully cooled disks in which strong magnetic braking regulates the spectral evolution without generating an outflow.

The observed optical–UV decay power-law indices provide a diagnostic not only of fallback dynamics but also of the thermal and angular momentum transport properties of the accretion disk. In our framework, these effects are parametrized by $\epsilon_{\rm rad}$, the ratio of radiative cooling to viscous heating, and $\psi$, which characterizes the strength of magnetic braking. A wind is launched when $\epsilon_{\rm rad} < 1$, as surplus heat is available to drive mechanical outflows. The wind mass loss is further enhanced as $\psi$ increases, due to more efficient angular momentum extraction by magnetic braking. As a result, the observed light-curve behavior encodes information about both the thermodynamic and dynamical state of the disk.

For example, TDEs whose light curves are well fitted by exponential decays—such as ASASSN-14ae and ASASSN-14li \citep{2014MNRAS.445.3263H,2016MNRAS.455.2918H} may reflect disks with efficient cooling or rapidly changing thermal states. In contrast, TDEs like ASASSN-18pg, which return to a $t^{-5/3}$ decay at late times \citep{2020ApJ...898..161H}, may correspond to long-lived, moderately cooled disks in which winds are continuously driven.

In summary, the observed clustering around $n_{\mathrm{OUV}} \approx 1.5$ is naturally reproduced in our models when either radiative cooling is inefficient ($\epsilon_{\rm rad} \ll 1$) or magnetic torques are sufficiently strong ($\psi \gtrsim 1$) to enable wind launching. As shown in Figure~\ref{fig:OUV_slope_obs}, even no-wind disks with $\epsilon_{\rm rad} = 1$ can match the observed power-law indices if magnetic braking is strong ($\psi = 10$), whereas the case with $\psi = 0$ fails to do so until very late times. In this sense, the observed decay behavior appears broadly consistent with a picture in which disk-driven outflows and their associated thermal and dynamical regulation play an important role in shaping the optical–UV emission. Within the framework of our self-similar disk solutions, this interpretation offers a unified explanation for the diversity of observed TDE light-curve power-law indices. Further testing with a broader range of theoretical models and statistical samples will be essential to assess its general applicability.

Our models reproduce the time evolution (i.e., the decay power-law index) of the integrated optical–UV luminosity $L_{\mathrm{OUV}}(t)$ quite well, but the absolute luminosity levels remain modest—typically no higher than $10^{42}$ erg/s. This is 1–2 orders of magnitude below the peak luminosities of many well-observed TDEs, such as ASASSN-14li and PS1-10jh, which can reach up to $10^{44}$ erg/s \cite{2012Natur.485..217G,Holoien2016,vanVelzen2020}. This discrepancy likely stems from our choice of relatively compact initial disk radii $r_0$ and surface densities $\Sigma_0$ (see equations~\ref{eq:rt} and \ref{eq:sigma0}  for specific parameters that determine these values). In observed TDEs, it is plausible that the disk extends to larger radii or that external irradiation (e.g., reprocessing of X-rays) enhances the observed flux. The primary aim of this study has been to clarify the temporal evolution of the light curve, particularly the decay power-law index, rather than to match the absolute luminosity. Addressing the latter will require more comprehensive models that incorporate radiative transfer and disk thermal structure in greater detail, which we defer to future work.


%
\subsection{Comparison with X-ray–selected TDEs}
%
While our primary focus in this study has been on the optical–UV band, it is worth considering whether the same disk–wind framework can also account for the temporal evolution observed in non-jetted, X-ray–bright TDEs. In such events, the emission is thought to originate from the innermost regions of the accretion flow, so that the observed X-ray flux is expected to track the bolometric luminosity, $L_{\rm bol}$, more directly than in the optical-UV band, as described in the previous subsection.

To this end, we adopt $n_{\rm l}$ as a first-order proxy for the X-ray power-law index $n_X(t)$. Our solutions with high cooling-to-heating flux ratio ($\epsilon_{\rm rad} \gtrsim 0.9$) and moderately strong magnetic braking ($\psi \gtrsim 1$) naturally yield $n_{\rm l}$ in the range $-1.8 \lesssim n_{\rm l} \lesssim -1.4$
\footnote{
Similar values of $n_{\rm l}$ also appear in other parameter regions, such as $\epsilon_{\rm rad} \sim 0.5$ with $\psi \gtrsim 3$. However, X-ray emission is expected to trace the bolometric luminosity more directly when radiative cooling is efficient and the inner disk contributes significantly. Therefore, we focus on high $\epsilon_{\rm rad}$ cases as a more representative regime for comparison with X-ray-selected TDEs.}, as shown in Figure~\ref{fig:slope_comp}, which is consistent with the decay power-law indices observed in several non-jetted X-ray TDEs, such as 2XMMi~J184725.1$-$631724 \citep{2011ApJ...738...52L}, ASASSN-14li \citep{2016MNRAS.455.2918H}, and AT2018fyk \citep{2019MNRAS.488.4816W}. A more rigorous, spectrally resolved comparison, based on coupling the present dynamical framework with dedicated radiative transfer calculations in the X-ray regime, is deferred to future work.

%
\section{Conclusions}
\label{sec:conclusion}
%

We have derived the self-similar solutions for a one-dimensional, time-dependent, geometrically thin accretion disk with a magnetically driven, non-relativistic outflow. In our model, we adopt the $\alpha$-parameter prescription to describe the MHD turbulent viscosity and magnetic braking. The model is characterized by three key parameters: $\bar{\alpha}_{r\phi}$, representing the turbulent viscosity; $\psi$, characterizing the magnetic braking; and $\epsilon_{\rm rad}$, denoting the ratio of radiative cooling to heating fluxes in the disk. We obtained two sets of self-similar solutions corresponding to the cases $(\beta_{+},\omega_{+})$ and $(\beta_{-},\omega_{-})$, where $\omega_\pm$ and $\beta_\pm$ are functions of $\epsilon_{\rm rad}$ and $\psi$ (see equations \ref{eq:omega} and \ref{eq:beta}). The solutions for $(\beta_{-},\omega_{-})$ predict an increase of the disk mass with time in the presence of the wind, leading to unphysical results which we exclude in this study. In contrast, the solutions for $(\beta_{+},\omega_{+})$ indicate that the disk mass decreases with time. In this case, the angular momentum is conserved in the absence of wind, but decreases in the presence of wind. Our main conclusions regarding the physically plausible solutions $( \beta_{+},~\omega_{+})$ are summarized as follows:

\begin{enumerate}
\item Our self-similar solutions show that the mass accretion rate, wind mass loss rate, and bolometric luminosity evolve as power laws in time. The bolometric luminosity deviates from the accretion rate at early to middle times when $\epsilon_{\rm rad} \ll 1$ and $0\le\psi \lesssim 1$, but it follows the accretion rate at very late times regardless of the values of $\epsilon_{\rm rad}$ and $\psi$. The power-law indices depend explicitly on $\epsilon_{\rm rad}$ and $\psi$: both the accretion and mass loss rates decay more steeply with decreasing $\epsilon_{\rm rad}$ and increasing $\psi$.

\item 
We confirm that, in the absence of the wind, the self-similar solution reduces to the classical solution of Cannizzo et al. (1990) \cite{1990ApJ...351...38C}: both the mass accretion rate and the bolometric luminosity follow the power law $t^{-19/16}$.
\item 
When the wind is present ($0<\epsilon_{\rm rad}<1$ and $\psi\ge0$), the mass accretion and loss rates decay more steeply with time than $t^{-19/16}$. 
Additionally, the mass accretion rate has a steeper power-law decay in time than the mass loss rate.
\item
We confirm that the time power-law indices of the mass accretion and loss rates match the numerical solutions provided by Tamilan et al. (2024) \cite{2024ApJ...975...94T} at late times, implying that the numerical solutions in this regime asymptotically approach the self-similar solutions.
\item 
In the limit $\psi \rightarrow \infty$, where magnetic braking dominates disk evolution, the time power-law indices of the mass accretion rate, mass loss rate, and bolometric luminosity asymptote to $-5/2$, independent of the value of $\epsilon_{\rm rad}$. This steep index may serve as evidence of magnetocentrifugally driven winds with strong poloidal $B$-fields in the context of TDEs.
\end{enumerate}

\section*{Acknowledgment}
We thank the referee for the constructive suggestions that have improved the paper.

\section*{Funding}
M.T. and K.H. have been supported by the Basic Science Research Program through the National Research Foundation of Korea (NRF) funded by the Ministry of Education (2016R1A5A1013277 to K.H. and 2020R1A2C1007219 to K.H. and M.T.). This work is also supported by Grant-inAid for Scientific Research from the MEXT/JSPS of Japan, 22H01263 to T.K.S. This research was supported in part by grant no. NSF PHY-2309135 to the Kavli Institute for Theoretical Physics (KITP).

\appendix

%
\section{Temporal evolution of the disk quantities for the case of $\omega = \omega_-$}
\label{app:msolution}
%

In this section, we present the time evolution of the disk mass, angular momentum, mass accretion rate, mass loss rate, and the bolometric luminosity for the case of $\omega = \omega_-$.

\subsection{Disk mass and angular momentum}

Using equation (\ref{eq:md}), the disk mass is given by
\begin{equation}
	M_{\rm d} 
	=
	4\pi\Sigma_0r_0^2
	\left[
	\frac{4}{15}
	\omega
	\left(
	1+
	\frac{1}{5}
	\left[
	\psi + 7
	-\frac{\beta}{\omega}
	\right]
	\right)^{-1}
	\right]^{3/2}
	G_1(\xi_{\rm in})
	\,
	\tau^{-\beta + 4 \omega}, 
\end{equation} 
where $G_1(\xi_{\rm in})$ is given by equation (\ref{eq:G1t}). By approximating the disk mass as $M_{\rm d} \propto \tau^{n_{\rm M}}$, the power-law index $n_{\rm M}$ is then given by 
\begin{equation}\label{app:nM}
n_{\rm M} = \frac{\partial \ln G_1(\xi_{\rm in}(\tau))}{\partial \ln \tau}-\beta+4\omega .
\end{equation}
 Equation~(\ref{eq:G1split}) for the $\omega = \omega_-$ case yields $G_1(\xi_{\rm in}) \propto \xi_{\rm in}^{(-1 - 3\psi + 3\beta/\omega)/5}$ as $\xi_{\rm in} \rightarrow 0$ when
 \begin{equation}
     \psi > \frac{50 + 27\epsilon_{\rm rad}}{30 - 9\epsilon_{\rm rad}},
 \end{equation}
and approaches a finite steady value otherwise. This divergence is clearly illustrated in Figure~\ref{fig:mdjd}(b): for $\psi = 10$ (dashed curves), $G_1(\xi_{\rm in})$ grows sharply at small $\xi_{\rm in}$, especially at lower $\epsilon_{\rm rad}$ values. Therefore, the power-law index $n_{\rm M}$ is then given by 
\begin{equation}\label{app:nMv}
	n_{\rm M}	=  \left\{
	\begin{array}{ll}
		\displaystyle{\frac{3}{2}\left[\frac{\sqrt{(1+\psi)^2+8(1-\epsilon_{\rm rad})(3+\psi)}-\psi-1}{\psi+15-\sqrt{(1+\psi)^2+8(1-\epsilon_{\rm rad})(3+\psi)}}\right]} & \displaystyle{,\psi < \frac{50 + 27\epsilon_{\rm rad}}{30 - 9\epsilon_{\rm rad}} }\\
		\\
		\displaystyle{\frac{3}{5}\left[\frac{4\sqrt{(1+\psi)^2+8(1-\epsilon_{\rm rad})(3+\psi)}-\psi-15}{\psi+15-\sqrt{(1+\psi)^2+8(1-\epsilon_{\rm rad})(3+\psi)}}\right]} & \displaystyle{,\psi > \frac{50 + 27\epsilon_{\rm rad}}{30 - 9\epsilon_{\rm rad}}},
	\end{array}
	\right. 
\end{equation}
where we have used $\xi_{\rm in} = \sqrt{r_{\rm in}/r_0} \tau^{-\omega}$ and $\omega=\omega_-$ and $\beta=\beta_-$ from equations (\ref{eq:omega}) and (\ref{eq:beta}), respectively. In the absence of the wind, $\psi = 0$ and $\epsilon_{\rm rad} = 1$, $n_{\rm M} = 0$ indicating the disk mass is constant, consistent with Pringle (1991) \cite{1991MNRAS.248..754P}. In the limit $\psi \rightarrow \infty$, $n_{\rm M}$ approaches $\infty$.

Using equation (\ref{eq:jd}), the total angular momentum of the disk is given by
\begin{equation}
	J_{\rm d} 
	=
	4\pi {r_{0}^2} \Sigma_0 ( r_0^2 \Omega_0)
	\left[
	\frac{4}{15}
	\omega
	\left(
	1+
	\frac{1}{5}
	\left[
	\psi + 7
	-\frac{\beta}{\omega}
	\right]
	\right)^{-1}
	\right]^{3/2}
	G_2(\xi_{\rm in})
	\,
	\tau^{-\beta + 5 \omega}, 
\end{equation}
where $G_2(\xi_{\rm in})$ is given by equation (\ref{eq:G2t}). By approximating the time evolution of the disk angular momentum as $J_{\rm d} \propto \tau^{n_{\rm J}}$, the power-law index $n_{\rm J}$ is then given by 
\begin{equation}\label{app:nJ}
	n_{\rm J} = \frac{\partial \ln G_2(\xi_{\rm in}(\tau))}{\partial \ln \tau}-\beta+5\omega .
\end{equation}
Equation~(\ref{eq:G2split}) for the $\omega = \omega_-$ case yields $G_2(\xi_{\rm in}) \propto \xi_{\rm in}^{(4 - 3\psi + 3\beta/\omega)/5}$ as $\xi_{\rm in} \rightarrow 0$ for
\begin{equation}
   \psi > \frac{250 - 54\epsilon_{\rm rad}}{75 - 18\epsilon_{\rm rad}}, 
\end{equation}
and approaches a finite steady value otherwise. This divergence is clearly shown in Figure~\ref{fig:mdjd}(d): for $\psi = 10$ (dashed curves), $G_1(\xi_{\rm in})$ grows sharply at small $\xi_{\rm in}$. 
Therefore, $n_{\rm J}$ is then given by 
\begin{equation}\label{app:nJv}
	n_{\rm J}	=  \left\{
	\begin{array}{ll}
		\displaystyle{\frac{3}{2}\left[\frac{\sqrt{(1+\psi)^2+8(1-\epsilon_{\rm rad})(3+\psi)}-\psi+1}{\psi+15-\sqrt{(1+\psi)^2+8(1-\epsilon_{\rm rad})(3+\psi)}}\right]} & \displaystyle{,\psi < \frac{250 - 54\epsilon_{\rm rad}}{(75 - 18\epsilon_{\rm rad})} }\\
		\\
		\displaystyle{\frac{3}{5}\left[\frac{4\sqrt{(1+\psi)^2+8(1-\epsilon_{\rm rad})(3+\psi)}-\psi-15}{\psi+15-\sqrt{(1+\psi)^2+8(1-\epsilon_{\rm rad})(3+\psi)}}\right]} & \displaystyle{,\psi > \frac{250 - 54\epsilon_{\rm rad}}{(75 - 18\epsilon_{\rm rad})}},
	\end{array}
	\right. 
\end{equation}
where we have used $\xi_{\rm in} = \sqrt{r_{\rm in}/r_0} \tau^{-\omega}$ and $\omega=\omega_-$ and $\beta=\beta_-$ from equations (\ref{eq:omega}) and (\ref{eq:beta}), respectively. 

\begin{figure}
	\centering
	\subfigure[]{\includegraphics[scale = 0.45]{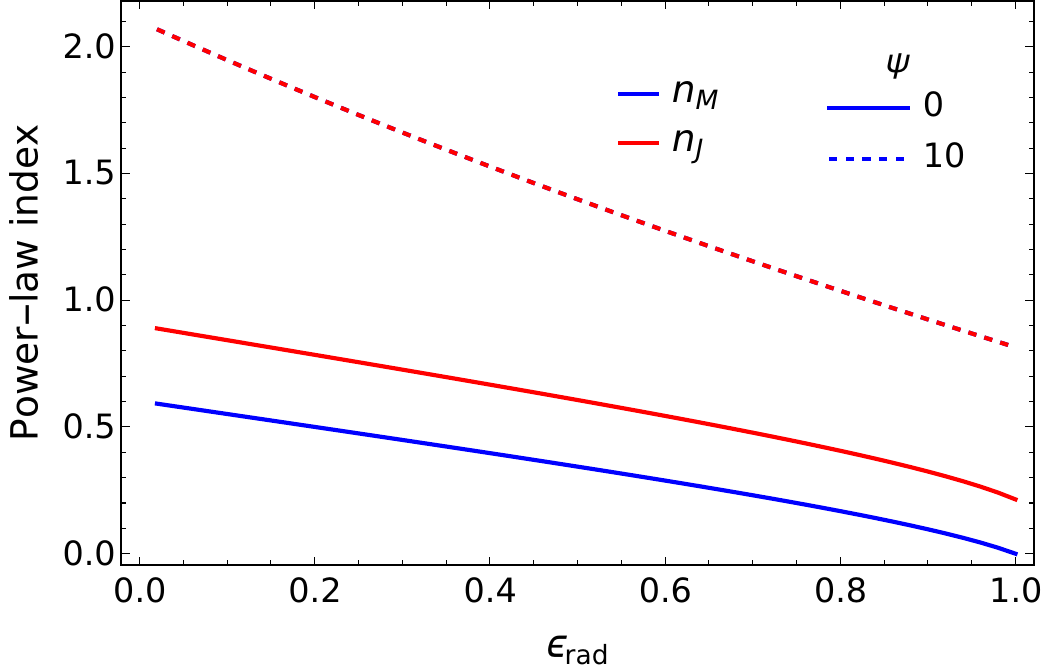}}
	\subfigure[]{\includegraphics[scale = 0.45]{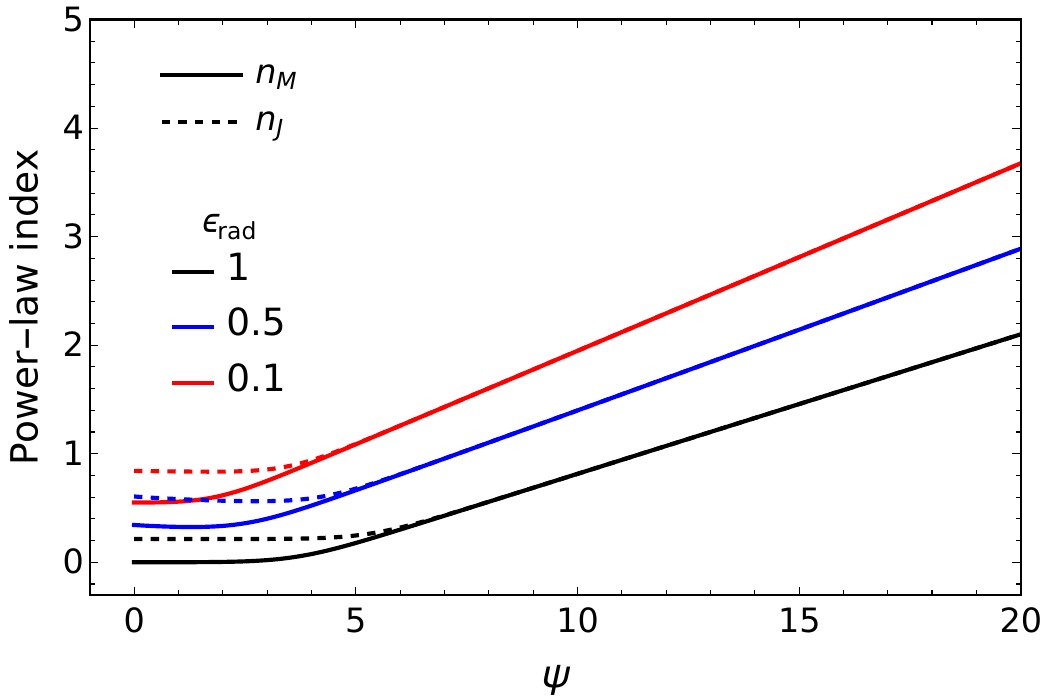}}
	\caption{The power-law indices for the time evolution of the disk mass ($n_{\rm M}$) and angular momentum ($n_{\rm J}$) are shown for the case $\omega = \omega_-$. Panel (a) presents $n_{\rm M}$ and $n_{\rm J}$ as functions of $\epsilon_{\rm rad}$ for two values of $\psi = 0$ and $10$. For $\psi = 0$, the indices differ, while for $\psi = 10$, they coincide. Panel (b) displays the dependence of these indices on $\psi$ for three values of $\epsilon_{\rm rad} = 0.1$, $0.5$, and $1$. At large $\psi$, the disk mass and angular momentum evolve with the same power-law index.}
	\label{fig:nMnJ_minus} 
\end{figure}

Figure~\ref{fig:nMnJ_minus} shows the power-law indices $n_{\rm M}$ and $n_{\rm J}$ for the $\omega = \omega_-$ solution. While $n_{\rm M}$ and $n_{\rm J}$ remain distinct at $\psi = 0$, they converge at large $\psi$. For $\epsilon_{\rm rad} = 1$ and $\psi = 0$, corresponding to a wind-free disk, the indices are $n_{\rm M} = 0$ and $n_{\rm J} = 3/14$, which indicate a constant disk mass and an increase in angular momentum driven by the outward expansion of the disk’s outer radius. This behavior aligns with the solution found by Pringle (1991) \cite{1991MNRAS.248..754P}. When $\epsilon_{\rm rad} < 1$ with $\psi \geq 0$, winds are present, and both $n_{\rm M}$ and $n_{\rm J}$ become positive, increasing further as $\epsilon_{\rm rad}$ decreases or $\psi$ increases. This implies a net growth of disk mass and angular momentum over time. However, such a growth cannot occur without external torque or mass supply, which we have not included in our model. Therefore, this behavior is physically counterintuitive, as both accretion and disk winds are expected to deplete the disk, not replenish it. We thus conclude that the $\omega = \omega_-$ solution is incompatible with realistic boundary conditions and do not adopt it in our main analysis.


%
%

\subsection{Mass accretion and loss rates}

Using equation (\ref{eq:mdacc2}), the mass accretion rate is given by
\begin{eqnarray}
	\dot{M}
	&=&
	3\pi
	\left[\frac{4}{15}\omega\left(1+\frac{1}{5}\left[\psi + 7-\frac{\beta}{\omega}\right]\right)^{-1}\right]^{5/2}
	\frac{\Sigma_0{r_0^2}}{t_0}
	\left(
	\frac{	r}{r_0} 
	\right)^{-(\psi+5-\beta/\omega)/2}
	\nonumber \\
	&\times&
	\biggr[\left(\frac{\beta}{\omega}-4\right)-(8+\psi)\xi^{
			2+\frac{2}{5}
			\left(
			\psi + 7 - \frac{\beta}{\omega}
			\right)}
		\biggr]
	\left[ 
	1 
	- 
	\xi^{2+\frac{2}{5}
		\left(
		\psi + 7 - \frac{\beta}{\omega}
		\right)} 
	\right]^{3/2}
	\tau^{\omega(\psi+7)-8\beta/3}.
\end{eqnarray}
Using equations (\ref{eq:omega}) and (\ref{eq:beta}) for $\omega_-$ and $\beta_-$, respectively, $\beta/\omega - 4 =$ \\ $ \left(1+\psi-\sqrt{(1+\psi)^2+8(1-\epsilon_{\rm rad})(3+\psi)}\right)/2 $. In the absence of wind, $\epsilon_{\rm rad} = 1$ and $\psi = 0$, this expression reduces to $\beta/\omega - 4 = 0$. However, in the presence of wind ($\epsilon_{\rm rad} < 1$ and $\psi \geq 0$), we find $\beta/\omega - 4 < 0$. This indicates that the mass accretion rate becomes negative for all values of $\epsilon_{\rm rad}$ and $\psi$ at every $\xi$, implying a radial outward flow of disk gas throughout the disk. By approximating $\dot{M} \propto \tau^{n_{\rm acc}}$, the power-law index $n_{\rm acc}$ is given by 
\begin{equation}\label{eq:nacc_minus}
	n_{\rm acc} = \frac{4\sqrt{(1+\psi)^2+8(1-\epsilon_{\rm rad})(3+\psi)}-\psi-15}{\psi+15-\sqrt{(1+\psi)^2+8(1-\epsilon_{\rm rad})(3+\psi)}}.
\end{equation}
In the absence of the wind, $\epsilon_{\rm rad} = 1$ and $\psi=0$, $n_{\rm acc} = -11/14$, which agrees with Pringle (1991) \cite{1991MNRAS.248..754P}.

Using equation (\ref{eq:mdw}), the mass loss rate is given by
\begin{equation}
	\dot{M}_{\rm w}=
	18 \pi 
	\left(1 - \epsilon_{\rm rad}\right)
	\left(1+\frac{\psi}{3}\right)
	\left[\frac{4}{15}\omega\left(1+\frac{1}{5}\left[\psi + 7-\frac{\beta}{\omega}\right]\right)^{-1}\right]^{5/2}
	\frac{\Sigma_0 r_0^2}{t_0} 
	 \tau^{2\omega-\frac{5}{3}\beta}
	\,
	G_3(\xi_{\rm in}),
\end{equation}
where $G_3(\xi_{\rm in})$ is given by equation (\ref{eq:G3t}). By approximating the mass loss rate as $\dot{M}_{\rm w} \propto \tau^{n_{\rm w}}$, the power-law index $n_{\rm w}$ is given by 
\begin{equation}\label{app:nw}
n_{\rm w} = \frac{\partial \ln G_3(\xi_{\rm in}(\tau))}{\partial \ln \tau} + 2\omega-\frac{5}{3}\beta.
\end{equation}
In equation (\ref{eq:G3split}), for $\omega = \omega_-$, the exponent $(\psi + 5 - \beta/\omega)$ evaluates to \\ $\left[\sqrt{(1+\psi)^2 + 8(1 - \epsilon_{\rm rad})(3 + \psi)} + \psi + 1\right]/2$, which is positive. This implies that $G_3(\xi_{\rm in}) \propto \xi_{\rm in}^{-(\psi + 5 - \beta/\omega)}$ diverges as $\xi_{\rm in} \rightarrow 0$. Consequently, the mass loss rate $\dot{M}_{\rm w}$ also diverges in this limit, indicating that a finite inner truncation is required to obtain a physically reasonable mass loss rate. Substituting $G_3(\xi_{\rm in}) \propto \xi_{\rm in}^{-(\psi + 5 - \beta/\omega)}$ and $\xi_{\rm in} = \sqrt{r_{\rm in}/r_0}\, \tau^{-\omega}$ into equation~(\ref{app:nw}) leads to 
\begin{equation}\label{eq:nw_minus}
	n_{\rm w} = \frac{4\sqrt{(1+\psi)^2+8(1-\epsilon_{\rm rad})(3+\psi)}-\psi-15}{\psi+15-\sqrt{(1+\psi)^2+8(1-\epsilon_{\rm rad})(3+\psi)}},
\end{equation}
which is identical to the power-law index of the mass accretion rate $n_{\rm acc}$ given by equation (\ref{eq:nacc_minus}).



\subsection{Disk luminosity}

The bolometric luminosity is given by 
\begin{eqnarray}
	L_{\rm bol}=\frac{32}{15}\pi^6 
	\frac{h r_0^2}{c^2}
	\left(\frac{k_{\rm B}T_{\rm eff,0}}{h}\right)^4
	\tau^{-5\beta/3} 
	\,
	G_4(\xi_{\rm in}), 
\end{eqnarray}
where $G_4(\xi_{\rm in})$ is given by equation (\ref{eq:G4t}). Assuming that $L_{\rm bol}\propto \tau^{n_{\rm l}}$, we get
\begin{eqnarray}
\label{app:nlt}
n_{\rm l} = \frac{\partial \ln{G_4(\xi_{\rm in}(\tau))}}{\partial \ln{\tau}}-\frac{5}{3}\beta.
\end{eqnarray}
Considering equation (\ref{eq:G4split}) in the case $\omega = \omega_{-}$, the exponent $(\psi + 7 - \beta/\omega)$ simplifies to $(\psi + 5 + \sqrt{(1 + \psi)^2 + 8(1 - \epsilon_{\rm rad})(3 + \psi)})/2$, which is strictly positive for $\epsilon_{\rm rad} \leq 1$ and $\psi \geq 0$. As a result, $G_4(\xi_{\rm in}) \propto \xi_{\rm in}^{-(\psi + 7 - \beta/\omega)}$ diverges as $\xi_{\rm in} \rightarrow 0$. This implies that the luminosity diverges as $r_{\rm in} \rightarrow 0$, so a finite inner truncation is required to obtain a physically reasonable luminosity. By substituting $G_4(\xi_{\rm in}) \propto \xi_{\rm in}^{-(\psi + 7 - \beta/\omega)}$ and $\xi_{\rm in} = \sqrt{r_{\rm in}/r_0} \,\tau^{-\omega}$ into equation (\ref{app:nlt}), the power-law index $n_{\rm l}$ is given by
\begin{equation}
	n_{\rm l} = \frac{4\sqrt{(1 + \psi)^2 + 8(1 - \epsilon_{\rm rad})(3 + \psi)} - \psi - 15}{\psi + 15 - \sqrt{(1 + \psi)^2 + 8(1 - \epsilon_{\rm rad})(3 + \psi)}},
\end{equation}
which matches the power-law indices of both the mass accretion rate $n_{\rm acc}$ in equation (\ref{eq:nacc_minus}) and the mass loss rate $n_{\rm w}$ in equation (\ref{eq:nw_minus}).

%
\begin{figure}
	\centering
	\subfigure[]{\includegraphics[scale = 0.45]{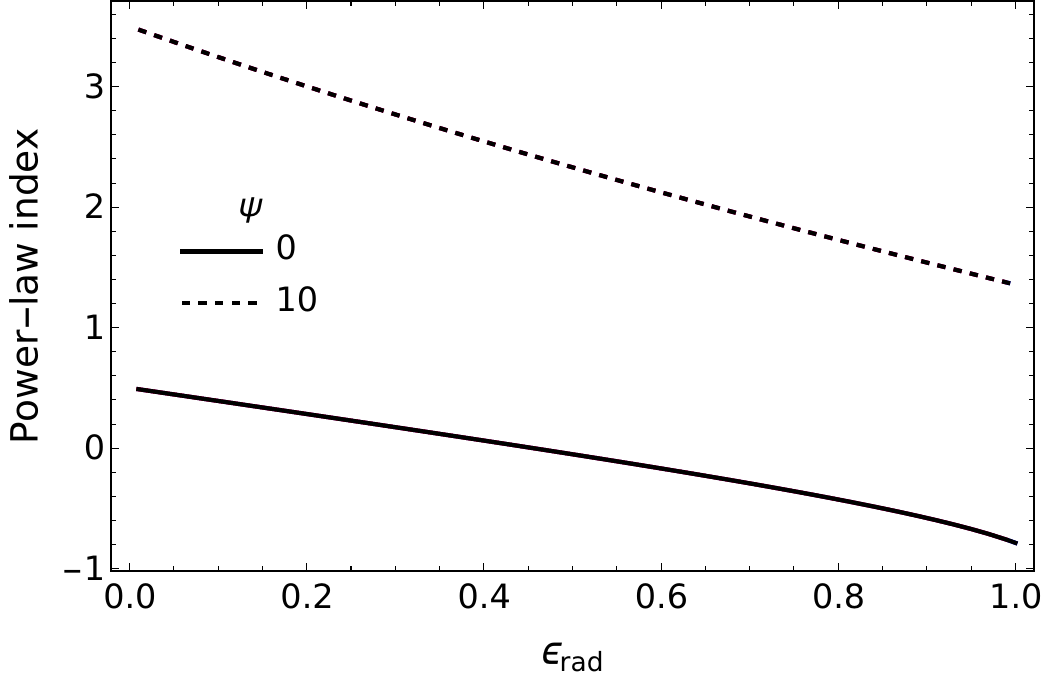}}
	\subfigure[]{\includegraphics[scale = 0.45]{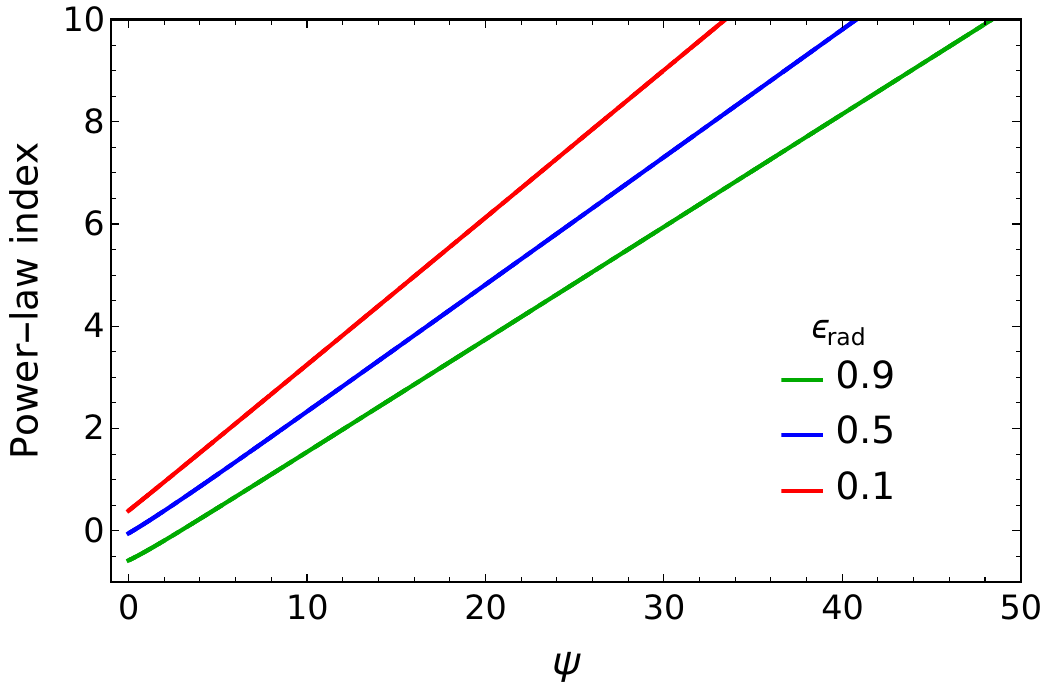}}
	\caption{\label{fig:nawl_minus} 
	The power-law indices for the time evolution of the mass accretion rate ($n_{\rm acc}$), wind mass loss rate ($n_{\rm w}$), and bolometric luminosity ($n_{\rm l}$) are shown for the case of $\omega = \omega_-$. Notably, $n_{\rm acc}$, $n_{\rm w}$, and $n_{\rm l}$ are found to be identical. Panel (a) displays the power-law indices as functions of $\epsilon_{\rm rad}$ for two representative values of $\psi = 0$ and $10$. Panel (b) illustrates the dependence on $\psi$ for three different values of $\epsilon_{\rm rad} = 0.1$, $0.5$, and $0.9$.}
\end{figure}

Figure \ref{fig:nawl_minus} shows the power-law indices for the time evolution of the mass accretion rate ($n_{\rm acc}$), wind mass loss rate ($n_{\rm w}$), and bolometric luminosity ($n_{\rm l}$). These indices are identical, indicating a shared temporal scaling for accretion, mass loss, and radiation output in the $\omega_-$ solution. Notably, the power-law indices become positive for low $\epsilon_{\rm rad}$ and high $\psi$, implying that the accretion rate, wind mass loss rate, and bolometric luminosity increase with time.

Although the $\omega = \omega_-$ solution is mathematically consistent, it exhibits several features that are incompatible with physically realistic boundary conditions. In particular, it predicts a net growth of the total disk mass and angular momentum over time when the wind is present ($\epsilon_{\rm rad}<1$ and $\psi \geq 0$), as indicated by the positive values of the power-law indices $n_{\rm M}$ and $n_{\rm J}$ in Figure \ref{fig:nMnJ_minus}. This implies that the disk is gaining mass and angular momentum without any external source, which is unphysical, as our model does not include an external mass supply or torque. Furthermore, Figure~\ref{fig:nawl_minus} shows that the mass accretion rate, wind mass loss rate, and bolometric luminosity all share a common power-law index of time that becomes positive for low $\epsilon_{\rm rad}$ and high $\psi$, indicating that all three quantities increase with time. In particular, the continued rise in luminosity is inconsistent with the expected evolution of a dissipative disk system, which naturally cools and loses energy as mass is accreted and lost through winds. In the absence of external heating, the luminosity of such a system should decline over time, reflecting the gradual depletion of available mass and internal energy. We therefore conclude that the $\omega = \omega_-$ solution does not describe a physically plausible scenario and exclude it from our main analysis, focusing instead on the $\omega = \omega_+$ solution.


%


%

\vspace{0.2cm}
\noindent

\let\doi\relax
\bibliographystyle{ptephy}
\bibliography{reference}

\end{document}